\documentclass[12pt,ta4paper]{article}

\usepackage[colorlinks,bookmarksopen,bookmarksnumbered,citecolor=red,urlcolor=red]{hyperref}

\usepackage{amsmath,amssymb}
\usepackage{textcomp} 
\usepackage{url}
\usepackage{natbib}
\usepackage[pdftex]{graphicx}

\usepackage{a4wide}
\usepackage[T1]{fontenc} 
\usepackage[utf8]{inputenc} 
\usepackage{tgschola}   

\newcommand\BibTeX{{\rmfamily B\kern-.05em \textsc{i\kern-.025em b}\kern-.08em
T\kern-.1667em\lower.7ex\hbox{E}\kern-.125emX}}

\newcommand{\fejer}{Fej\'er }

\title{A nestable, multigrid-friendly grid on a sphere for global spectral models based on Clenshaw-Curtis quadrature}

\author{Daisuke Hotta\thanks{Meteorological Research Institute, Japan Meteorological Agency, Tsukuba, Japan}
  \and Masashi Ujiie\thanks{Numerical Prediction Division, Japan Meteorological Agency, Tokyo, Japan}}

\begin{document}
\maketitle

 This is the peer reviewed version of the following article: Hotta and Ujiie (2018) {\it QJRMS}{\bf 144(714)} 1382--1397, which has been published in final form at \\ \url{https://doi.org/10.1002/qj.3282}. This article may be used for non-commercial purposes in accordance with Wiley Terms and Conditions for Self-Archiving.

\begin{abstract}
 A new grid system on a sphere is proposed that allows for
 straightforward implementation of both spherical-harmonics-based
 spectral methods and gridpoint-based multigrid methods. The latitudinal
 gridpoints in the new grid are equidistant and spectral transforms in
 the latitudinal direction are performed using Clenshaw-Curtis
 quadrature. The spectral transforms with this new grid and quadrature
 are shown to be exact within the machine precision provided that the
 grid truncation is such that there are at least $2N+1$ latitudinal
 gridpoints for the total truncation wavenumber of $N$. The new grid and
 quadrature is implemented and tested on a shallow-water equations model
 and the hydrostatic dry dynamical core of the global NWP model
 JMA-GSM. The integration results obtained with the new quadrature are
 shown to be almost identical to those obtained with the conventional
 Gaussian quadrature on Gaussian grid. Only minor code changes are
 required to adapt any Gaussian-based spectral models to employ the
 proposed quadrature.
\end{abstract}


\maketitle


\section{Introduction}
Global spectral atmospheric models that are in use today almost
universally adopt Gaussian quadrature in the meridional direction to
perform forward (gridpoint-to-wavenumber) spherical harmonics
transform. Gaussian quadrature is an optimal quadrature rule in the
sense of maximising the degree of polynomials that can be integrated
exactly for a given number of quadrature points (or nodes). Given $J$
nodes, Gaussian quadrature is exact for integrand polynomials of up to
as high as $2J-1$ degrees. This optimality is achieved, however, with
several inconveniences \citep[e.g.][]{cc60}: First, the nodes and
weights are not given in an explicit analytic form and necessitates
(some iterative) solution of an algebraic equation of high
degrees. Second, the nodes {\it do not nest}, i.e., the nodes for the
$J$-point rule do not contain any $J'$-point nodes as their subset for
any $J' < J$. While the former is now not so much an inconvenience than
it used to be thanks to the recently developed elegant, fast and yet
accurate computing methods \citep[e.g.,][and the other algorithms
reviewed in Townsend, 2015]{Hale-Townsend-13}, all the more so in the
context of atmospheric modelling since the quadrature nodes and weights
can be pre-computed once at the initialisation process and stored in
memory to be reused during the subsequent time integration steps, the latter
limitation does impose some inflexibility to the future evolution of a
global spectral dynamical core. In particular, the unnestable grid
alignment makes it cumbersome, if not impracticable, to combine the current
spectral dynamical core with a multigrid approach for solving an
elliptic boundary value problem that arises from (semi-)implicit time
discretisation.

As the horizontal resolution continues to increase, it becomes necessary
to properly represent nature's non-hydrostatic aspects in the model. In
a non-hydrostatic system, a semi-implicit time stepping results in a
Helmholtz problem with spatially-variable coefficients that needs to be
solved iteratively, even with horizontal spectral discretisation
\citep[e.g.,][]{benard04}. This is in contrast to the hydrostatic case
where the resultant Helmholtz problem has constant coefficients and thus
can be solved without iteration in the spectral space \citep{hs75}.

The multigrid approach is an attractive strategy for iterative solution
of the non-hydrostatic implicit Helmholtz problem. It accelerates
convergence of the iterative algorithm by first solving the problem at a
lower resolution and then gradually increasing the resolution, ingesting
the solution from the previous (lower) resolution as the initial guess
to the next (higher) resolution. While such an approach has been applied and
proved effective in the context of grid-based atmospheric models
\citep[e.g.][]{hrk13,stvd15}, implementing it on current spectral models
is not straightforward because the aforementioned non-nested nature
of Gaussian quadrature necessitates some form of accurate off-grid
interpolation \citep[e.g.,][]{jones99,ullrich09} from higher- to
lower-resolution grid (and vice versa). See Appendix C for further discussion.

 Another spectrally accurate family of quadrature rules exist, however,
that, unlike the conventional Gaussian quadrature, have nestable nodes
that would allow for straightforward implementation of multigrid
approach. These quadrature rules, introduced by \cite{fejer33} and
\cite{cc60}, have been well known in the field of Numerical Analysis and
are shown by some authors to have several advantages over the classical
Gaussian quadrature. Although Clenshaw-Curtis or \fejer quadrature rules
are not optimal in the sense of giving exact integration for polynomials
of up to only $J-1$ (as opposed to the Gaussian $2J-1$) degrees given
$J$ nodes, they have been shown to be practically as accurate as
Gaussian quadrature in many applications \citep[e.g.][and the references
therein]{trefethen08}.  Despite gaining popularity in numerical
analysis, to the authors' best knowledge, Clenshaw-Curtis-type
quadrature appears not to have been used in spectral transform models,
at least in atmospheric modelling.

In this paper we aim to show that the Clenshaw-Curtis quadrature, with
its associated nodes (which turn out to be just equispaced latitude
grids, see (\ref{eq:cc-points})), can be used as an alternative to the
classical Gaussian quadrature with the Gaussian latitude grids in
atmospheric global spectral models. Although our ultimate goal is to
investigate the effectiveness of a multigrid approach in a global
non-hydrostatic spectral model defined on the multigrid-friendly
Clenshaw-Curtis grid, we defer actual multigrid implementation to future
work and limit the scope of the present paper to only showing numerical
soundness of the quadrature applied to the associated Legendre functions
and the equivalence of the Gaussian and Clenshaw-Curtis quadrature
implemented on a shallow-water equations (SWE) model and a
three-dimensional dry hydrostatic primitive equations (HPE) model.  To
clarify our motivation, we provide, in Appendix C, a rough outline on
how the proposed quadrature and grid will aid in multigrid
implementation.

The rest of the paper is structured as follows: Section 2 reminds the
reader with how the quadrature is used in global spectral atmospheric
models. Section 3 introduces Clenshaw-Curtis quadrature in the context
of spectral transform. Section 4 examines the numerical orthonormality
of the associated Legendre functions evaluated with Clenshaw-Curtis
quadrature with a comment on aliasing of quadratic and cubic
terms. Sections 5 and 6 document the implementation of Clenshaw-Curtis
quadrature to the spherical SWE model and the HPE global dynamical core,
and compare the results produced with Gaussian and Clenshaw-Curtis
quadrature in the context of standardised test cases. Section 7
concludes the paper with an outlook for our future plans.

\section{Discrete spherical harmonics transforms and Gaussian quadrature}

Global spectral models represent atmospheric state variables in both
gridpoint space and wavenumber (spectral) space and transform them back
and forth during the course of time integration \citep{orszag70}.

The discrete inverse spherical harmonics transform, or synthesis,
transforms the state variable from the spectral representation $X_n^m$
to its gridpoint-space representation $X(\lambda_i, \phi_j)$ by the following
formula:
\begin{eqnarray}
 X(\lambda_i,\phi_j)
  &=& \sum_{m=-N}^N\sum_{n=|m|}^{N} X_n^m Y_n^m(\lambda_i,\phi_j) \label{eq:iSHT}
\end{eqnarray}
where $N$ is the truncation total wavenumber, $n$ and $m$ are,
respectively, the total and zonal wavenumber, $\lambda$ is the
longitude, $\phi$ is the latitude, $Y_n^m(\lambda,\phi)$ is the
spherical harmonic function with the total and zonal wavenumbers of $n$
and $m$, respectively. Throughout this manuscript, the triangular
truncation as in (\ref{eq:iSHT}) is assumed.  In the Japan
Meteorological Agency (JMA)'s Global Spectral Model (JMA-GSM), as in
other spectral models, the transform of (\ref{eq:iSHT}) is implemented
via two steps by first performing the associated Legendre transform in
the meridional direction from $X_n^m$ to
\begin{eqnarray}
 X^m(\phi_j):=\sum_{n=|m|}^{N}X_n^m\tilde{P}_n^m(\sin\phi_j) \label{eq:iLT}
\end{eqnarray}
and then performing inverse discrete Fourier transform (DFT) in the
zonal direction from $X^m(\phi_j)$ to
$X(\lambda_i,\phi_j):=\sum_{m=-N}^{N}
X^m(\phi_j)e^{\sqrt{-1}\lambda_i}$. Here, $\tilde{P}_n^m(\sin\phi_j)$ denotes the
associated Legendre polynomial of degree $n$ and order $m$, normalised
to satisfy the following orthonormality:
\begin{eqnarray}
 \int_{-\frac{\pi}{2}}^{\frac{\pi}{2}}\tilde{P}_n^m(\sin\phi)\tilde{P}_{n'}^m(\sin\phi)\cos(\phi)d\phi=\delta_{n,n'}. \label{eq:pnm-ortho}
\end{eqnarray}
The longitudinal gridpoints $\{\lambda_i\},i=0,\cdots,I-1$ are chosen so that
the DFT can be computed efficiently via a Fast Fourier Transform (FFT) algorithm,
resulting in equispaced grids $\lambda_i=2\pi i/I$.

The direct spherical harmonics transform, or analysis, recovers
spectral coefficients $X_n^m$ from the gridpoint-space representation
$X(\lambda_i,\phi_j)$, by exploiting the orthonormality of the spherical harmonics:
\begin{eqnarray}
 X_n^m &=&\int_{-\frac{\pi}{2}}^{\frac{\pi}{2}}\int_{0}^{2\pi}X(\lambda,\phi)
  {Y_n^{m}}^*(\lambda,\phi)d\lambda \cos{\phi}d\phi \label{eq:dSHT} \\  
 &=& \int_{-1}^{1}
  X^m(\phi)\tilde{P}_n^m(\sin\phi) 
  d\left(\sin\phi\right) \label{eq:dLT}
\end{eqnarray}
with
  \begin{eqnarray}
   X^m(\phi)=\int_{0}^{2\pi}X(\lambda,\phi)e^{-\sqrt{-1}m\lambda}d\lambda  \label{eq:dFT}
  \end{eqnarray}
The integrations in (\ref{eq:dLT}-\ref{eq:dFT}) have to be
evaluated numerically with some quadrature rules. For the zonal
direction, direct DFT gives exact integration to an integrand with wavenumbers up to $I$
given $I$ zonal grids. Thus, for each meridional gridpoint
$\phi_j$, (\ref{eq:dFT}) can be discretised as
\begin{eqnarray}
 X^m(\phi_j)&=&\frac{1}{I}\sum_{i=0}^{I-1}X(\lambda_i,\phi_j)e^{-\sqrt{-1}m\lambda_i}
\end{eqnarray}
with the constraint 
\begin{eqnarray}
 I > 2N \label{eq:fourier-condition}
\end{eqnarray}
to avoid aliasing introduced by quadrature error (note that the
integrand of (\ref{eq:dFT}) contains zonal wavenumbers of at most $2N$ because
$X(\lambda,\phi)$ is a truncated sum of $Y_n^m$ with $|m|$ being at most $N$).
For the meridional direction, any quadrature rule can be used as long as it gives exact
integration to polynomials of $\sin\phi$ to a desired degree. In
practice, however, Gaussian quadrature (with the ordinary Legendre
polynomials as the basis set) is almost always employed due to its
optimality in the sense described in the first paragraph of Section 1.
With Gaussian quadrature, (\ref{eq:dLT}) is discretised as
\begin{eqnarray}
 X_n^m &=& \sum_{j=1}^{J} X^m(\phi_j) \tilde{P}_n^m(\sin\phi_j) w^G_j  \label{eq:gauss-quad}
\end{eqnarray}
with the Gaussian weights given by
\begin{eqnarray}
 w^G_j &=& \frac{2}{JP'_J(\sin\phi_j)P_{J-1}(\sin\phi_j)} \label{eq:gauss-wgt}
\end{eqnarray}
and the latitude grid (the Gaussian latitudes) $\{\phi_j\},j=1,\cdots,J$
given as the $J$ roots of $P_J(\sin\phi)=0$ where $P_J(\sin\phi)$ is the
ordinary Legendre polynomial of degree $J$ \citep{DLMF}.  We remark that
several exquisitely fast and accurate algorithms for computing the
Gaussian latitudes $\phi_j$ and weights $w^G_j$ have been recently
worked out by numerical analysts. Unlike the classical methods employed
in atmospheric models that are based on Newton iteration
\citep[e.g.,][]{Swartztrauber02,enomoto15} or tridiagonal eigen-solution
\citep{AS97}, the recently developed methods exploit very accurate
asymptotic expansions on $P_n(\sin\phi)$ and other related functions and
prove particularly useful for high-resolution modelling with large
$J$. Notably, \cite{bogaert14} successfully derived asymptotic
expressions for $\phi_j$ and $w^G_j$ that are accurate enough to allow
iteration-free evaluation of them with errors below double-precision
machine epsilon; see \cite{Townsend15} for a historical review.

In order to avoid aliasing due to quadrature errors, there should be at
least
\begin{eqnarray}
 J\geq (2N+1)/2 \label{eq:gauss-condition}
\end{eqnarray}
latitudinal gridpoints for the given truncation total wavenumber $N$ . This is
because the integrand of (\ref{eq:dLT}) is a polynomial of $\sin\phi$ of
at most $2N$ degree\footnote{Note that, in spectral methods,
  $X^m(\phi)$ is assumed to be a linear
 combination of $\left\{\tilde{P}_{n'}^m(\sin\phi)\right\} (|m|\leq n'\leq
 N) $  and that $\tilde{P}_{n}^m(\sin\phi)$ takes the form of
 $\left(1-\sin^2\phi\right)^{|m|/2}\times\left(\mbox{polynomial of\
 }\sin\phi \mbox{\ of degree at most\ } n-m \right)$.}.
This condition, along with
(\ref{eq:fourier-condition}), allows exact discrete transforms with  linear truncation (i.e., $J\approx N$).

\section{Clenshaw-Curtis quadrature}

In numerically evaluating the integration in (\ref{eq:dLT}), it is
possible, though apparently never have been tried with global spectral
models, to use a quadrature rule other than the classical Gaussian
formula (\ref{eq:gauss-quad}--\ref{eq:gauss-wgt}). In this paper we
focus on using one variant of Clenshaw-Curtis-type quadrature in place
of the standard Gaussian quadrature, motivated by its nestable property
that we described in Introduction. The quadrature rule we use, whose
nodes (\ref{eq:cc-points}) do not contain the poles, should probably be
more appropriately called \fejer quadrature of second kind
\citep{fejer33,waldvogel06,trefethen08}. Following \cite{boyd87},
however, in this manuscript we shall abusively call it Clenshaw-Curtis quadrature.

\subsection{Formulation}
\label{sec:cc-formulation}
For convenience, we denote the integrand of (\ref{eq:dLT})  by
\begin{eqnarray}
 g(\cos\theta):=f(\sin\phi):=X^m(\phi)\tilde{P}_n^m(\sin\phi) \label{eq:defg}
\end{eqnarray}
where we have introduced the colatitude
$\theta=\frac{\pi}{2}-\phi$. Clenshaw-Curtis-type quadrature rules first
expand $g(\cos\theta)$ into some trigonometric series and then
analytically integrate them term-by-term. The choice of trigonometric
base functions specifies the quadrature points and the weights. In
atmospheric modelling, more specifically in transitioning from a
spherical harmonics transform model based on Gaussian quadrature, it is
desirable to avoid placing quadrature points on the poles (c.f., section
\ref{sec:md72}). We therefore choose to expand $g(\cos\theta)$ into
Chebyshev polynomials of the second kind
$U_l(\cos\theta)=\sin{(l+1)\theta}/\sin{\theta}$ of degrees
$l=0,1,\cdots,J-1$, or equivalently, to expand $g(\cos\theta)\sin\theta$
into the sine series $\sum_{p=1}^J a_p\sin{p\theta}$, which can be
achieved by performing the type-I discrete sine transform (DST) on
$g(\cos\theta_j)\sin\theta_j$ without introducing any errors but
rounding. From this constraint, the collocation points $\theta_j$ (or
$\phi_j$) are chosen as
\begin{eqnarray}
 \theta_j=\frac{j}{J+1}\pi,
  \phi_j=\frac{\pi}{2}\left(1-\frac{2j}{J+1}\right), j=1,2,\cdots,J \label{eq:cc-points}
\end{eqnarray}
Note here that the poles ($j=0$ and $j=J+1$) are absent, which is
exactly what motivated us to choose $U_l(\cos\theta)$ as the base
functions.

As pointed out by Gentleman (1972a,1972b), the type-I DST in expanding
$g(\cos\theta)$ into Chebyshev polynomials $U_l(\cos{\theta})$ can be efficiently
implemented by FFT, allowing quadrature to be performed with
$O(J\log{J})$ operations without explicit computation of
the quadrature weights. While this is advantageous if a single execution
of quadrature is concerned, this is not the case for spherical-harmonics-based global spectral
models where the quadrature of the form of (\ref{eq:dLT}) is repeated
$O(N^2)$ times on each spectral transform and the weights can be
pre-computed and stored in memory to be reused many times. From the
viewpoint of adapting the code from the existing one based on Gaussian
quadrature, it is favourable if the quadrature can be expressed in the
standard interpolative form analogous to (\ref{eq:gauss-quad}) rather
than to rely on FFT (or DST) code. \cite{boyd87} showed that this is
possible. He showed that the quadrature can be cast in the following
standard quadrature form:
\begin{eqnarray}
 \int_{-1}^{1}f(\sin\phi)d\left(\sin\phi\right)&=&\sum_{j=1}^{J}f(\sin\phi_j)w^{CC}_j \\
 &=& \sum_{j=1}^{J}X^m(\phi_j)\tilde{P}_n^m(\sin\phi_j)w^{CC}_j  \label{eq:cc-quad}
\end{eqnarray}
with the weights $w^{CC}_j$ given by
\begin{eqnarray}
 w^{CC}_j=\frac{4\sin\theta_j}{J+1}\sum_{\substack{1\leq p\leq J\\
 p\mathrm{:\ odd}}}\frac{\sin\left(p\theta_j\right)}{p}. \label{eq:cc-wgt}
\end{eqnarray}
\cite{boyd87} did not provide explicit derivation of (\ref{eq:cc-wgt});
we give a concise derivation in Appendix A.  Computing the
Clenshaw-Curtis weights from (\ref{eq:cc-wgt}) involves $O(J^2)$
operations. This is usually not a significant burden for atmospheric
models since the weights can be computed only once at the model's
initialisation and then can be stored on memory, but we remark that a faster
$O(J\log{J})$ algorithm for computing $w^{CC}_j$ exploiting FFT has been 
devised by \citep[][their Eqs. (3.5, 3.9, 3.10)]{waldvogel06}\footnote{Note
that the quadrature rule we use in the present paper is referred to as
\fejer's second rule in this paper.}.

Equation (\ref{eq:cc-quad}) and its Gaussian counterpart
(\ref{eq:gauss-quad}) take the same form, and the Clenshaw-Curtis
latitudes and weights are symmetric across the Equator just like their
Gaussian equivalent, which means that only minor code adaptation is
necessary to implement Clenshaw-Curtis quadrature on any model based on
Gaussian quadrature. For example, in the case of JMA-GSM, code
adaptation took addition of only $\sim$ 20 lines.  One deviation that
might affect implementation is that, unlike the Gaussian grids,
Clenshaw-Curtis grids include the Equator; its implication on code
adaptation and hemispheric symmetry of solutions is discussed in
Appendix B.

An important difference from Gaussian quadrature is that the number of nodes, given the
truncation total wavenumber $N$, required for alias-free exact 
meridional integration, is
\begin{eqnarray}
 J\geq 2N+1 \label{eq:cc-condition},
\end{eqnarray}
which is about twice larger than in the Gaussian case
(\ref{eq:gauss-condition}).  This is because Clenshaw-Curtis quadrature
expands the integrand into Chebyshev polynomials of up to $(J-1)$-th
degree and thus $J-1$ needs to be no less than $2N$, the maximum degree
of the integrand polynomial (see the footnote below
(\ref{eq:gauss-condition})). 

Conventionally, in the context of spectral modelling, truncation rules
with $J\approx 2N$, $J\approx \frac{3}{2}N$ and $J\approx N$ are
referred to, respectively, as ``cubic,'' ``quadratic'' and ``linear''
truncation, which are so named because the quadratic truncation, for
instance, avoids aliasing when a quadratic term computed in the grid
space is transformed back to spectral space (and likewise for cubic and
linear truncation). With this nomenclature, the condition
(\ref{eq:cc-condition}) requires Clenshaw-Curtis quadrature to be used
with ``cubic'' (or higher-order) truncation to assure exact spectral
transforms on linear terms. Strictly speaking, it is perhaps an abuse of
terminology to apply this nomenclature to Clenshaw-Curtis quadrature
since, for example, ``linear'' truncation does not guarantee alias-free
transform on linear terms. In this manuscript, we nevertheless adhere to
this convention since, as we show in section \ref{sec:cc-aliasing},
aliasing errors that arise from inexact quadrature are orders of
magnitude smaller than those that arise from sub-sampling in grid space
and hence pose little problem in practice.

It may seem that having to use cubic truncation is too
much a restriction for Clenshaw-Curtis quadrature to be usable in an
atmospheric model. Recent findings in the context of very
high-resolution atmospheric simulation suggest, however, that this is
perhaps not really a serious restriction: \cite{wedi14} for instance
showed, through experimentation using the spectral atmospheric model of
the European Centre for Medium-Range Weather Forecast (ECMWF), that the
aliasing errors coming from the quadratic and cubic (or even
higher-order) terms on the right-hand-side of the governing equations
get larger as the horizontal resolution increases, so that,
at a resolution as high as $\sim$ 10 km grid spacing, cubic truncation
(e.g., Tc1023), which automatically filters out quadratic and cubic
aliasing, yields more accurate forecasts than the linear
truncation with the same grid spacing (e.g., Tl2047) does. ECMWF in fact
adopted cubic truncation at the total wavenumber of 1279 ($\sim 8$ km
grid spacing) into their operational suite in 2016
\citep{malardel16}. It is thus likely that future high-resolution global
spectral models adopt cubic or higher-order truncation, in which case
the condition (\ref{eq:cc-condition}) is not a particular disadvantage.

\subsection{Lower resolution spherical harmonics transforms at a reduced cost}
\label{sec:lowres-sht}
The advantage of using Clenshaw-Curtis quadrature is that the quadrature
points nest. This not only allows for straightforward multigrid
implementation in gridpoint space but also allows spherical
harmonics transforms with a lower truncation wavenumber to be performed
economically. Suppose, for instance, the model has Tc479 (triangular
cubic truncation at total wavenumber 479) horizontal resolution with
Clenshaw-Curtis grid, and one wishes to compute spectral coefficients
of total wavenumber up to 239 (the lower half of the spectral space)
from the physical data on the Tc479 grid. For convenience let
$N_\mathrm{full}+1=480$, $N_\mathrm{half}+1=240$ and assume that, for a
truncation wavenumber $N$, there are $J(N)=2N+1$ latitudinal
gridpoints. Then, the colatitudes $\theta_j$ of Tc479 grid are
\begin{eqnarray}
 \theta_j=
  \frac{j\pi}{J(N_\mathrm{full})+1}=\frac{j\pi}{2\left(N_\mathrm{full}+1\right)}
\end{eqnarray}
with $j=1,2,\cdots,2N_\mathrm{full}+1$. 
If we take every other latitudinal gridpoints starting from $j=2$, the
colatitudes of this subset are
\begin{eqnarray}
 \theta_{2j'}
  =\frac{2j'\pi}{2\left(N_\mathrm{full}+1\right)}
  =\frac{j'\pi}{2\left(N_\mathrm{half}+1\right)}
  =\frac{j'\pi}{J(N_\mathrm{half})+1}
\end{eqnarray}
with $j'=1,2,\cdots,2N_\mathrm{half}+1$, which are identical to the
colatitudes of Tc239 Clenshaw-Curtis grids. The same nested property
also applies to the equispaced longitudinal grid as long as the number
of longitudinal gridpoints is chosen to be multiple of two. Thus, with
Clenshaw-Curtis quadrature, Tc479 grid contains Tc239 grid as its
complete subset, enabling us to compute spectral coefficients of the total
wavenumber up to 239 using direct (grid-to-wave) transform code of Tc239
model. How these mechanisms will help to simplify multigrid implementation is discussed in Appendix C.

\subsection{Discrete spherical harmonics transform on a shifted equispaced latitude grid}

In deriving the quadrature rule in section \ref{sec:cc-formulation}, we
chose to expand $g(\cos\theta)$ into the Chebyshev polynomials of the
second kind $U_l(\cos\theta)$. Alternatively, we could also expand
$g(\cos\theta)$ into the Chebyshev polynomials of the first kind
$T_l(\cos\theta):=\cos{l\theta}$ (\fejer's first rule). Then, the expansion can be performed
exactly by type-II DCT, resulting in the quadrature nodes 
\begin{eqnarray}
 \theta_j=\frac{j-1/2}{J}\pi,
  \phi_j=\frac{\pi}{2}\left(1-\frac{2j-1}{J}\right), j=1,2,\cdots,J \label{eq:fejer1-points}
\end{eqnarray}
which, compared to the Clenshaw-Curtis grid (\ref{eq:cc-points}), are shifted by half a grid, and the weights are
\begin{eqnarray}
 w^{F1}_j=\frac{2}{J}\left(1-\sum_{p=1}^{J/2}\frac{\cos\left(2p\theta_j\right)}{4p^2-1}\right). \label{eq:fejer1-wgt}
\end{eqnarray}

The grid (\ref{eq:fejer1-points}) is not nestable and thus is not suitable
for combining a multigrid approach, but its equispaced nature may be
useful in some applications. For example, this grid is identical to the
one used by double-Fourier-series-based spectral models
\citep{cheong00,yoshimura12}, so comparison between such models and a
traditional spherical-harmonics-based model may be facilitated by
running the latter with this grid and quadrature. 
We have implemented this quadrature to the SWE model along with the
Clenshaw-Curtis quadrature and confirmed that the test case results, as
shown in section 5, are almost identical to those of the Clenshaw-Curtis
version (not shown).

\subsection{Comparison with the discrete spherical harmonics transform of \cite{md72}}

\label{sec:md72} We remark that a global spectral atmospheric model on a
nestable equispaced latitudinal grid has once been developed very early
by \cite[][hereafter referred to as MD72]{md72}. Their quadrature rule
is specifically designed for integration of the form (\ref{eq:dLT}) and
was later analysed in detail by \cite{Swartztrauber79}.  The latitudinal
grid used by MD72 quadrature is almost identical to our
Clenshaw-Curtis grid (\ref{eq:cc-points}) except that the former
includes the poles ($j=0$ and $J+1$). Unlike our Clenshaw-Curtis
quadrature, however, their quadrature gives exact transform given $J
\geq N$, allowing for linear truncation. This is because their
quadrature expands $X^m(\phi)$ itself, rather than
$X^m(\phi)\tilde{P}_n^m(\sin\phi)$, into cosine (for even $m$) or sine
(for odd $m$) series in $\theta$: the Fourier coefficient $X^m(\phi)$ is
a linear combination of trigonometric functions with wavenumbers up to
$N$, so exact expansion into cosine or sine series is possible with the
trapezoid rule given only $J\geq N$ nodes (compare this with the
discussion for Clenshaw-Curtis quadrature given above right after
(\ref{eq:cc-condition})). The resultant formula takes the form
\begin{eqnarray}
 X_n^m = \sideset{}{''}\sum_{j=0}^{J+1} \bar{Z}_{n,j}^m X^m(\phi_j),
\end{eqnarray}
where the double-primed sum ($\Sigma''$) indicates trapezoidal summation with
the first and last terms in the summand given the weight of 1/2, and the
``weights'' $\bar{Z}_{n,j}^m$, which are different for each pair of $(n,m)$,
are defined using the integrals of the form
$\int_{-1}^{1}\sin{k\phi}\tilde{P}_n^m(\sin\phi)d\left(\sin\phi\right)$ and
$\int_{-1}^{1} \cos{k\phi}\tilde{P}_n^m(\sin\phi)d\left(\sin\phi\right),\ k=0,\cdots,N$
\citep[for details of derivation, see Section 4
of][]{Swartztrauber79}, to be evaluated exactly by Gaussian quadrature using more
than $(2N+1)/2$ Gaussian nodes. Computation of the ``weights''
$\bar{Z}_{n,j}^m$ by direct evaluation of their definitions given in
MD72 or \cite{Swartztrauber79} requires $O(N^4)$ operations, which would
be impracticable for high-resolution modelling, but \cite{AS97} later
devised an efficient $O(N^3)$ algorithm which uses a recurrence relation
for $\bar{Z}_{n,j}^m$ which in turn derives from the ``four-point
recurrence'' \citep[Eq. (5.12) of][]{AS97} for
$\tilde{P}_n^m(\sin\phi)$ .

Compared to Clenshaw-Curtis quadrature, MD72's quadrature is
advantageous in that it allows for linear truncation to be used, albeit
at the cost of increased amount of pre-computation to prepare the
``weights'' $\bar{Z}_{n,j}^m$. A rather serious drawback of MD72's
quadrature is that the nodes contain the poles: many modern global
spectral atmospheric models, including those of JMA, ECMWF and National
Centers for Environmental Prediction (NCEP), employ the ``$U$-$V$''
formulation \citep{ritchie88,temperton91} to represent horizontal vector
fields, which requires division of Fourier coefficients by $\cos^2\phi$
in conversions between the vector representation and the
rotation-divergence representation. Having the poles (for which
$\cos\phi$ is zero) thus induces division by zeroes, which is a trouble
from numerical point of view. The Clenshaw-Curtis grid without the poles
neatly avoids this problem. MD72's quadrature could be modified to avoid
the poles by employing the shifted grid (\ref{eq:fejer1-points}),
allowing the sine or cosine expansions to be exactly performed with
type-II DCT or type-II DST, but such a modification results in loss of
nestability.

\section{Numerical accuracy of spectral transforms based on Clenshaw-Curtis quadrature}
\subsection{Numerical orthonormality of the associated Legendre functions}
\label{sec:orthonormality}

Spectral transform methods \citep{orszag70} transform spectral and
gridpoint space back and forth on each time step by using the direct
and inverse spherical harmonics transforms. Because the transforms are
repeated many times during the course of time integration, discrete transforms
need to be nearly exact (i.e., the errors must be within the range of
rounding errors). In particular, the orthonormality of the associated
Legendre functions, (\ref{eq:pnm-ortho}), with the integral evaluated by
a quadrature rule, needs to strictly hold for all $m$ and $n$ within the
truncation limit. In the previous section we have seen that, in theory,
the strict quadrature should be guaranteed by imposing the condition
(\ref{eq:cc-condition}). To verify the {\it numerical} exactness on an
actual implementation, this subsection examines the numerical orthonormality
of the associated Legendre functions (\ref{eq:pnm-ortho}) evaluated with
Clenshaw-Curtis quadrature implemented on JMA-GSM. Here, we define {\it
normality error} and {\it orthogonality error}, respectively, as:
\begin{eqnarray}
 {\varepsilon_{n,\mathrm{N}}^m} &:=& 
  \left| 
   \sum_{j=1}^{J} \tilde{P}_n^m\left(\sin\phi_j\right)^2 w_j -1
  \right|    \label{eq:norm-err} \\
 {\varepsilon_{n,\mathrm{O}}^m} &:=&
  \max_{\substack{n'\neq n \\ m\leq n'\leq N}}
  \left|
   \sum_{j=1}^{J} \tilde{P}_n^m\left(\sin\phi_j\right)\tilde{P}_{n'}^m\left(\sin\phi_j\right)w_j\right| \label{eq:orth-err}
\end{eqnarray}
where $\phi_j$ and $w_j$ are the nodes and weights defined for each
quadrature rule. We examine ${\varepsilon_{n,\mathrm{N}}^m}$ and
${\varepsilon_{n,\mathrm{O}}^m}$ for different combinations of
quadrature and truncation rules. Clenshaw-Curtis quadrature requires
cubic or higher-order truncation (\ref{eq:cc-condition}). We thus focus
mainly on cubic truncation with $J=2N+1$ meridional gridpoints for a
given truncation wavenumber $N$. Given the observations from the
literature \citep[e.g.,][]{trefethen08}, that Clenshaw-Curtis quadrature
is nearly as accurate as the Gaussian in many practical applications, it
is still interesting to look at linear ($J\approx N$) and quadratic
($J\approx 3N/2$) truncation.

Throughout this section, the discrete Legendre transforms are computed
using the code of JMA-GSM \citep{miyamoto06}. The normalised associated
Legendre functions $\tilde{P}_n^m\left(\sin\phi_j\right)$ are computed
by the three-point recurrence \citep[Eqs. (5--9) of ][]{enomoto15} with
quadruple precision arithmetic and then cast to double
precision. Computation of (\ref{eq:norm-err}) and (\ref{eq:orth-err}) is
performed using double precision.
JMA-GSM employs the reduced spectral transform \citep{miyamoto06,juang04} in
which the terms involving $\tilde{P}_n^m\left(\sin\phi_j\right)$ whose
absolute value is below $10^{-16}$ are neglected from the summations in
(\ref{eq:iLT}) and (\ref{eq:gauss-quad}). We have repeated all
computations with this option on or off and found that this option does
not affect the results shown in this section. Here we only show the
results obtained with this option switched on.

\begin{figure}[htbp]
\centering
 \includegraphics[angle=90,width=\linewidth,bb=47 0 650 631]{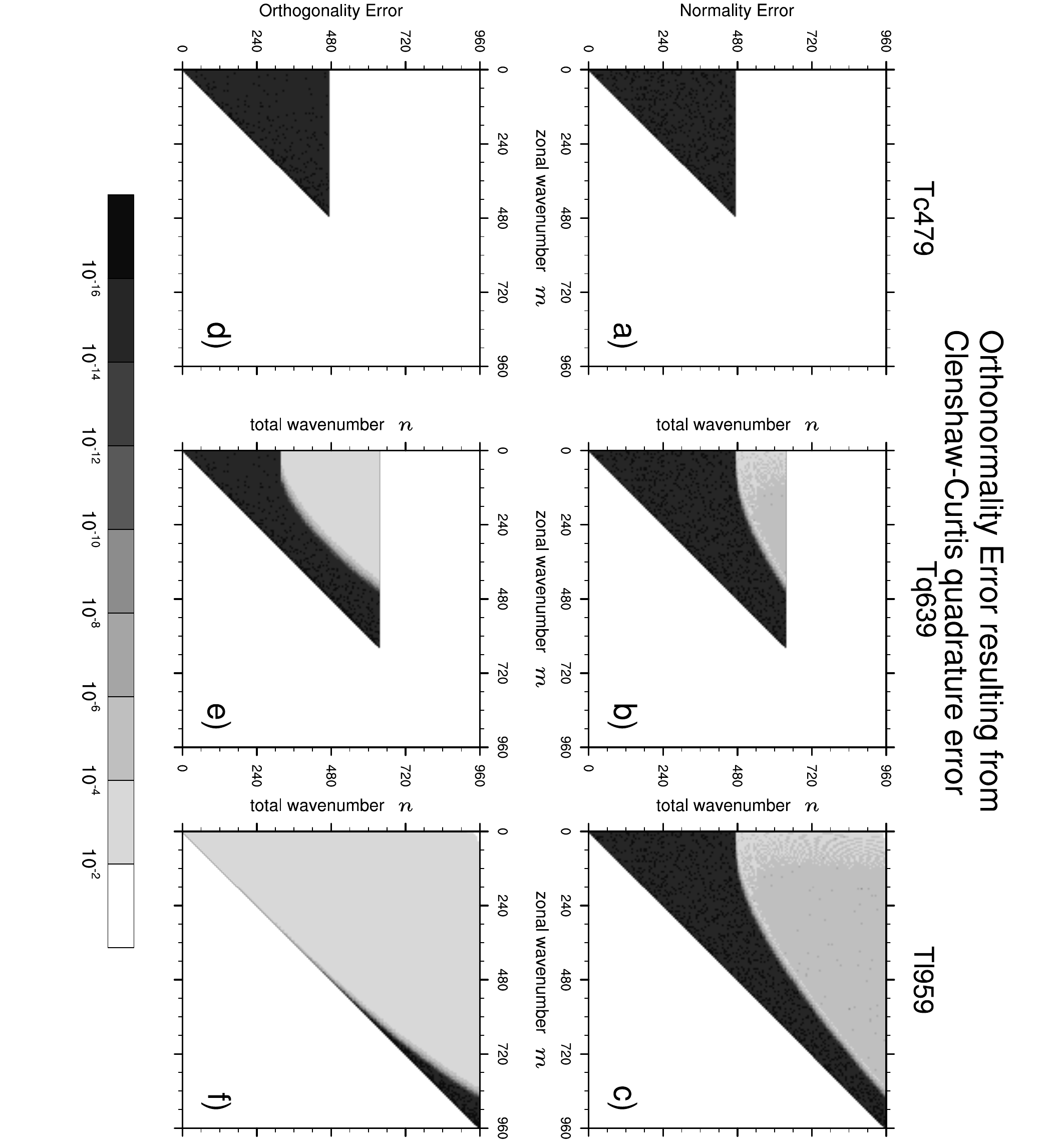}
 \caption{(a-c) Normality error (Eq. (\ref{eq:norm-err})) and (d-f)
 orthogonality error (Eq. (\ref{eq:orth-err})), evaluated using
 Clenshaw-Curtis quadrature, plotted as a function of the zonal
 wavenumber $m$ and the total wavenumber ($n$). The resolutions are
 (a,d): Tc479 ($N=479, J=959$), (b,e): Tq639 ($N=639, J=959$) and (c,f)
 Tl959 ($N=959, J=959$). Note the logarithmic scale in the colours.}
 \label{fig:orthonormality-959}
\end{figure}

 Figure \ref{fig:orthonormality-959} shows the normality error and
orthogonality error for Clenshaw-Curtis quadrature with the resolutions
of Tc479 (triangular cubic truncation at $N=479$ total wavenumber),
Tq639 (triangular quadratic truncation at $N=639$ total wavenumber) and
Tl959 (triangular linear truncation at $N=959$ total wavenumber), all
with $J=959$ meridional gridpoints. As the theory predicts, with cubic
truncation, both normality error (panel a) and orthogonality error
(panel d) are within the range of double-precision rounding ($\leq
10^{-16}$) for all combinations of $m$ and $n$. In contrast, with
lower-order truncation (panels b and c), normality error exceeds the
level of rounding error for total wavenumbers $n > (J-1)/2=479$, with a
tendency to become larger for larger $n$ and smaller $m$. This is
consistent with the condition (\ref{eq:cc-condition}) because the
integrand $\tilde{P}_n^m\left(\sin\phi\right)^2$ is a degree-$2n$
polynomial of $\cos{\theta}$ since it takes the form
$\sin^{2m}{\theta}\times\left[\mbox{polynomial of\ }\cos{\theta}\mbox{\
of degree\ } 2(n-m)\right]$
$=\left(1-\cos^2\theta\right)^m\times\left[\mbox{polynomial of\
}\cos{\theta}\mbox{\ of degree\ } 2(n-m)\right]$
$=\left[\mbox{polynomial of\ }\cos{\theta}\mbox{\ of degree\ }
2n\right]$. Interestingly, however, the normality error remains $<
10^{-3}$ even when the quadrature limit (\ref{eq:cc-condition}) is
violated (c.f., section \ref{sec:why-small}).
 
Clenshaw-Curtis quadrature guarantees orthogonality only to total
wavenumbers $ n\leq J-N-1$ because the degree of the integrand
$\tilde{P}_n^m\left(\sin\phi\right)\tilde{P}_{n'}^m\left(\sin\phi\right)$
in (\ref{eq:orth-err}), which is $n+n'\leq n+N$, need to be less than
$J$. Consistently, orthogonality is exact with cubic truncation (panel d) but not
with quadratic truncation (panel e) for the total wavenumbers $n\geq
J-N+1=320$. With linear truncation (panel f) it is not exact even for
$n=0$. Interestingly again, the orthogonality error nevertheless remains $<
10^{-3}$ for any pairs of $(n,m)$ that violate the quadrature limit (c.f., section \ref{sec:why-small}).

From these results we can conclude that Clenshaw-Curtis quadrature
assures exact spectral transforms if it is used with cubic (or
higher-order) truncation. This conclusion holds to all other resolutions
that we tested.

\subsection{Aliasing of quadratic and cubic terms evaluated on cubic Clenshaw-Curtis grids}
\label{sec:cc-aliasing}

Clenshaw-Curtis quadrature with cubic truncation does not guarantee
alias-free spectral transforms for quadratic and cubic terms. As such,
before applying Clenshaw-Curtis quadrature with cubic truncation to
nonlinear models, we need to quantitatively assess the degree of
aliasing for nonlinear terms. In this subsection we quantify
aliasing errors for quadratic and cubic terms associated with spherical
harmonics transforms and compare the Clenshaw-Curtis quadrature with
cubic truncation and the Gaussian quadrature with linear truncation.

We quantify the aliasing errors by the following procedure. Consider the
following series of operations:
\begin{enumerate}
 \item Produce three scalar fields $U$, $V$, and $W$ on the sphere by
       randomly choosing their spectral representations $U_n^m$, $V_n^m$
       and $W_n^m$.  The spectral coefficients are chosen such that
       their $(m,n)$-component, $0\leq n \leq N_\mathrm{trunc}=479$,
       $|m|\leq n$, has both its real and imaginary part taken as an
       independent pseudo-random draw from a uniform distribution over
       $[1\times (n+1)^{-1/3},2\times (n+1)^{-1/3}]$.
 \item Transform the scalar fields $U_n^m,V_n^m,W_n^m$ in spectral space
       into grid space by (\ref{eq:iSHT}) to form $U(\lambda_i,\phi_j)$,
       $V(\lambda_i,\phi_j)$, $W(\lambda_i,\phi_j)$, and then compute,
       in grid space, quadratic and cubic terms
       $X_q(\lambda_i,\phi_j):=U(\lambda_i,\phi_j)\times
       V(\lambda_i,\phi_j)$ and
       $X_c(\lambda_i,\phi_j):=U(\lambda_i,\phi_j)\times
       V(\lambda_i,\phi_j)\times W(\lambda_i,\phi_j)$.
 \item Finally transform the quadratic and cubic terms in grid space
       $X_q$ and $X_c$ back to spectral space by
       (\ref{eq:dLT}--\ref{eq:dFT}) to yield $X_{q,n}^m$ and
       $X_{c,n}^m$.
\end{enumerate}
The scaling of the pseudo-random numbers by $(n+1)^{-1/3}$ in the step 1
is intended to mimic the power spectra of physical quantities typically
encountered in geophysical applications (for instance, the power spectra
of kinetic energy in three-dimensional turbulence in the inertial range
follow $n^{-5/3}$-law, which corresponds to the spectral coefficients of
divergence or vorticity being proportional to $\sim n^{-1/3}$). This
choice is arbitrary and subjective, but we confirmed that it does not
sensitively affect the conclusion by repeating the calculation changing
the power from $-1/3$ to $-3,-2,-1,-1/2$ and 0.

We first perform the steps 1--3 on Tl1439 Gaussian grid ($N=1439,
J=1440$) using Gaussian quadrature and a linear truncation at the total
wavenumber of $N_\mathrm{ref}=1439 (\approx 3N_\mathrm{trunc})$. This
resolution is high enough to produce aliasing-free reference solutions
of both quadratic and cubic terms, which we denote respectively by
$X_{q,n}^{m,\mathrm{ref}}$ and $X_{c,n}^{m,\mathrm{ref}}$. We then
repeat the steps 1--3 using the test resolutions and quadrature rules
(Tl479 Gaussian or Tc479 Clenshaw-Curtis), using the same values of
$U_n^m$, $V_n^m$ and $W_n^m$ as used to produce the reference
solutions. The spectral coefficients of the quadratic and cubic terms
computed with the test resolution and quadrature rule, denoted
$X_{q,n}^{m,\mathrm{test}}$ and $X_{c,n}^{m,\mathrm{test}}$, are finally
compared with the reference solutions $X_{q,n}^{m,\mathrm{ref}}$ and
$X_{c,n}^{m,\mathrm{ref}}$ to define the normalised aliasing error for
each pair of $(n,m)$:
\begin{eqnarray}
 \mathrm{AliasingErr}_{q,n}^m &:=& {\left|X_{q,n}^{m,\mathrm{test}}-X_{q,n}^{m,\mathrm{ref}}\right|} / {\left|X_{q,n}^{m,\mathrm{ref}}\right|} \\
 \mathrm{AliasingErr}_{c,n}^m &:=& {\left|X_{c,n}^{m,\mathrm{test}}-X_{c,n}^{m,\mathrm{ref}}\right|} / {\left|X_{c,n}^{m,\mathrm{ref}}\right|}.
\end{eqnarray}

\begin{figure}[htbp]
\centering
 \includegraphics[width=0.8\linewidth,bb=47 0 578 631]{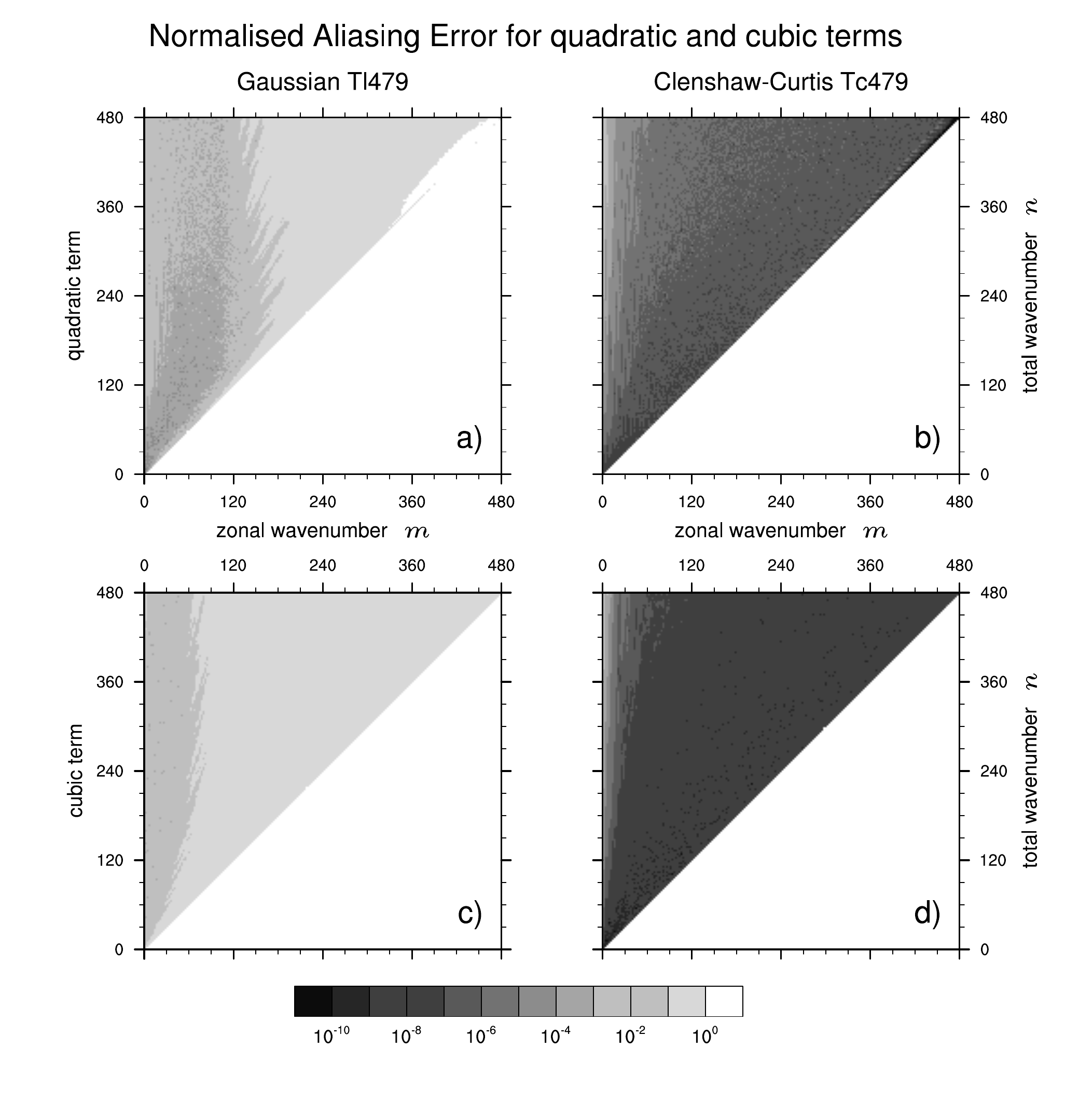}
 \caption{Normalised aliasing errors (see the text for definition) for
 (a,b) quadratic and (c,d) cubic terms evaluated with (a,c) Tl479
 Gaussian quadrature with linear grid ($N=479, J=480$) and (b,d) Tc479
 Clenshaw-Curtis quadrature with cubic grid ($N=479, J=959$). Note the
 logarithmic scale in the colours.}
 \label{fig:aliasing-err}
\end{figure}

The results for Gaussian linear grid (Tl479; $N=479, J=480$) and
Clenshaw-Curtis cubic grid (Tc479; $N=479, J=959$) are plotted in Figure
\ref{fig:aliasing-err}. Clenshaw-Curtis quadrature performed on cubic
grid (Figure \ref{fig:aliasing-err}, right panels) produces normalised aliasing
errors below $10^{-5}$ (in the quadratic case, Figure
\ref{fig:aliasing-err}b) or $10^{-8}$ (in the cubic case, Figure
\ref{fig:aliasing-err}d) for most pairs of $(n,m)$ except in a small
region with large $n/m$ ratio ($\gtrsim 10$). This is in contrast to the
case of Gaussian quadrature performed on linear grid (Figure
\ref{fig:aliasing-err}, left panel), where the normalised errors are almost
anywhere greater than $10^{-4}$ (in the quadratic case, Figure
\ref{fig:aliasing-err}a) or $10^{-3}$ (in the cubic case, Figure
\ref{fig:aliasing-err}c), for some pairs of $(n,m)$ even exceeding $1$
in the quadratic case.

These results indicate that the aliasing errors that arise from the
quadrature rule not being exact (as seen in Figure
\ref{fig:aliasing-err}b,d) are orders of magnitude smaller than the
aliasing errors that result from sub-sampling of waves in gridpoint
space (as seen in Figure \ref{fig:aliasing-err}a,c); we elaborate on
this in the next subsection. From this we may deduce that the aliasing
from nonlinear terms in a model with Clenshaw-Curtis quadrature should
remain under a controllable level as long as the quadrature is used with
cubic truncation. This conclusion is further corroborated by the results
of experiments shown in the next two sections.

\subsection{Explanation of small aliasing error following \cite{trefethen08}}
\label{sec:why-small}

In Figures \ref{fig:orthonormality-959} and \ref{fig:aliasing-err} we
have observed that Clenshaw-Curtis quadrature gives fairly small errors
even when the quadrature limit (\ref{eq:cc-condition}) is violated. This
interesting feature can be understood from the mechanism discussed in
section 5 of \cite{trefethen08} which provides an intuitive and rigorous
proof as to why the nominal factor-of-two advantage of
Gaussian quadrature over Clenshaw-Curtis is rarely realised
in practice. Trefethen's argument is based on the version of
Clenshaw-Curtis-type quadrature where the integrand function is expanded
into $T_l(\cos\theta)$, the Chebyshev polynomials of the first kind;
below we give a parallel discussion, expanding the integrand into
$U_l(\cos\theta)$ instead of $T_l(\cos\theta)$.

On the Clenshaw-Curtis grid (\ref{eq:cc-points}), it follows from the
$2\pi$-periodicity of the sine function and its antisymmetry around 0, that,
for any positive integer $p\leq J$, $U_{J+p}(\cos\theta)$ are
indistinguishable (or aliased onto) $U_{J-p}(\cos\theta)$ (with the sign
flipped):
\begin{eqnarray}
 U_{J+p}(\cos\theta_j) = - U_{J-p}(\cos\theta_j), j=1,2,\cdots,J,
\end{eqnarray}
from which follows that $I_J^{CC}(U_{J+p})=-I_J^{CC}(U_{J-p})=-I(U_{J-p})=0$ (if $J\pm p$ is
odd) or $-\frac{2}{J+1-p}$ (if $J\pm p$ is even), where
$I(g):=\int_{-1}^1g(\cos\theta)d(\cos\theta)$ and $I_J^{CC}(g)$ denotes
its approximation evaluated with $J$-point Clenshaw-Curtis quadrature
rule (note that $I_J^{CC}(U_{J-p})$ is exact since $I_J^{CC}(\cdot)$ is exact
for any polynomials with degrees at most $J-1$).  Consequently the
quadrature error for $U_{J+p}$ (which we denote by $e_J(p)$) is zero for
odd $J+p$ and
\begin{eqnarray}
 e_J(p):=I(U_{J+p})-I_J^{CC}(U_{J+p})=\frac{2}{J+1+p}+\frac{2}{J+1-p}
\end{eqnarray}
for even $J+p$. Now, assuming that $J$ is odd (as is the case for a
nestable grid), and expanding the integrand $g(\cos\theta)$ into
$\sum_{l=0}^\infty a_l U_l(\cos\theta)$ which we assume to be uniformly
converging, its quadrature error reads
$a_{J+1}e_J(1)+a_{J+3}e_J(3)+a_{J+5}e_J(5)+\cdots$. Since $e_J(p)\sim
O(J^{-1})$ for small $p$, and $a_{J+p}$ should rapidly decay as $p$
increases provided that $g(\cos\theta)$ is sufficiently smooth (which
should be a valid assumption for most geophysical applications), the
quadrature error for $g(\cos\theta)$ should be small if $J$ is large.

For example, with 959-point-Clenshaw-Curtis quadrature (Tc479
resolution), $U_{960}$ is integrated inexactly, but the error is only
$e_{959}(1)=0.004$; errors for $U_{1200},U_{1440},U_{1680}$ and
$U_{1920}$ are, respectively, 0.004, 0.005, 0.009 and 1.9.  Errors for
other polynomials are even smaller. For $T_{960},
T_{1200},T_{1440},T_{1680}$ and $T_{1920}$, the errors are 0.002,
2e$-$6, 8e$-$6, 3e$-$5 and 2.0; similarly, for Legendre polynomials
$P_{960}, P_{1200},P_{1440},P_{1680}$ and $P_{1920}$ whose exact
integrals are all zero, the errors are 8e$-$5, 4e$-$6, 5e$-$6, 1e$-$5
and 0.04. By contrast, with 479-point-Gaussian quadrature (Tl479 resolution),
$U_n, T_n$ or $P_n$ are all exactly evaluated up to $n=957$, but once
the polynomial degree goes beyond the quadrature limit
($n\leq 2J-1$), the errors jump from zero to 3.1, 1.6 and
0.06, respectively, for $U_{958}$, $T_{958}$ and $P_{958}$.

\section{Implementation to a shallow-water equation model}

In implementing a new scheme, it is convenient to test it with a simpler
model before introducing it to the full
three-dimensional  model. As a testbed, we first implemented
Clenshaw-Curtis quadrature to a semi-Lagrangian shallow-water equations
(SWE) model.  The main focus of this paper is to examine the consistency
between the models with Clenshaw-Curtis and Gaussian
quadrature. Accordingly, the validations that follow are not designed to
test the performance of the model itself but rather to detect any
discrepancies arising from the use of different quadrature rules.

\subsection{The model}

The model used in this section is built by adapting the code of JMA-GSM. The
governing equations are the advective form of the SWE on a
rotating sphere described in section 2.2 of \cite{williamson92} but with
the Coriolis terms incorporated in the left-hand side of the momentum equations as
in \cite{temperton97}. The governing equations in advective form are
then discretised in space and time by the Stable Extrapolation
Two-Time-Level Semi-Lagrangian (SETTLS) method of \cite{hortal02}. The
linear terms responsible for the fast gravity waves are treated
semi-implicitly using the second-order decentering method described in
Section 3.1.3 of \cite{yukimoto11}.  The semi-Lagrangian aspects of
the scheme, such as finding of, and interpolation to, the departure
points, are identical to the horizontal advection of JMA-GSM
\citep{yukimoto11,jma-outline13}. The reference state for the semi-implicit
linearisation is the fluid at rest with a globally constant depth (5960
m for \cite{williamson92} test and 10 km for \cite{galewsky04} test).
 Other parameters are chosen to conform to the specifications in
\cite{williamson92} and \cite{galewsky04} test cases. Throughout this
section, and for all resolutions, the time step is chosen as $\Delta t=1200$ s.

The model may include numerical (hyper-)diffusion term in the governing equation:
\begin{eqnarray}
 \left(\frac{\partial X}{\partial t}\right)_{\mathrm{diffusion}} = -K
  \nabla^{2r}X \label{eq:diffusion}
\end{eqnarray}
where $r$ is a positive integer. This is numerically solved
implicitly in spectral space by applying the following operation at
the end of each time step:
\begin{eqnarray}
 \left(X_n^m\right)_\mathrm{after} = \frac{1}{1+2K\Delta t
  \left\{\frac{n(n+1)}{a^2}\right\}^r} \left(X_n^m\right)_\mathrm{before}.
\end{eqnarray}
where $n$, $m$, $a$ and $\Delta t$ are, respectively, the total and
zonal wavenumbers, the Earth's radius and the time step.

We modify the model to allow an option to use Clenshaw-Curtis quadrature
and compare the model runs produced with this option and the runs
produced with the default Gaussian quadrature, under the framework of
two standard test cases. The models with the two different quadrature
rules run on different grids. To allow for accurate comparison,
the Clenshaw-Curtis version of the model is adapted to output model
states also on the Gaussian grids by converting the model state
variables in spectral space to grid space using the inverse spherical
harmonics transform (\ref{eq:iSHT}) defined on the Gaussian grids. To
ensure that the two versions start from exactly identical initial
conditions, the initial conditions for the Clenshaw-Curtis version are
produced by first computing them in grid space on the Gaussian grids,
converting them to spectral space by the direct spherical harmonics
transform (\ref{eq:dSHT}) defined on the Gaussian grids, and then
converting them back to grid space (but this time on the
Clenshaw-Curtis grids) by the inverse spherical harmonics transform
(\ref{eq:iSHT}) defined on the Clenshaw-Curtis grids.

\subsection{Test cases}

The first test case examined is the evolution of the initially zonal
flow over an isolated mountain proposed by \cite{williamson92} as Case
\#5. This test case is repeated for the resolutions of Tc31, Tc63,
Tc127 and Tc255, both for Clenshaw-Curtis and Gaussian version of the model. No
explicit numerical diffusion of the form (\ref{eq:diffusion}) is used in any
of the integrations for this case.

The solution of \cite{williamson92} test case \#5 appears to be
dominated by relatively low wavenumber components, so it may not be
suitable for assessing model behaviour at higher wavenumbers. The second
test case we examine, the barotropic jet instability test case proposed by
\cite{galewsky04}, allows us to assess the models in a more
realistic, multi-scale flow situations. \cite{galewsky04} reported that
weak explicit dissipation in the governing equations is necessary to
obtain converged solutions, and suggested to apply harmonic diffusion
($r=1$ and $K=1.0\times10^5~\mathrm{m}^2$ in (\ref{eq:diffusion})) to
the governing equations to facilitate model comparison. We followed this
suggestion and performed the test with a relatively high resolution of
Tc479 ($ \Delta x \sim$ 20 km at the Equator).

\subsection{Results: flow over an isolated mountain}
\label{sec:results-test5} 

In interpreting the results, it should be noted that, unlike models with
Eulerian advection scheme, exact agreement between Clenshaw-Curtis and
Gaussian versions of the model is not expected in our semi-Lagrangian
model since the two versions use different grids that result
in differences in departure point interpolation. Nevertheless, we found,
by plotting overlaid contour maps of relative vorticity (or any other)
fields from the two versions of the model (not shown), that the two
quadrature rules result in visually indistinguishable integrations up to
at least 288 hours, for any of the tested resolutions. To quantitatively
assess the difference, we computed the normalised $L_2$-differences defined as:
\begin{eqnarray}
 L_2\left(\xi^{CC},\xi^{G}\right):=\left(
 \frac{I\left(\xi^{CC}-\xi^{G}\right)}{I\left(\xi^{G}\right)}
\right)^{1/2}
\end{eqnarray}
for zonal wind ($\xi=u$), meridional wind ($\xi=v$) and height field
($\xi=h$), for integration times $t$ from 0 to 288 hours at 6-hourly intervals. Here the superscripts
$CC$ and $G$ represent the integration, respectively, by Clenshaw-Curtis
and Gaussian versions of the model, and the the spherical integral  $I(\xi)$ of the square of a variable $\xi$

\begin{eqnarray}
 I(\xi):=\frac{1}{4\pi}\int_{-\frac{\pi}{2}}^\frac{\pi}{2}\int_0^{2\pi} \xi(\lambda,\phi)^2 d\lambda d\cos{\phi}d\phi
\end{eqnarray}
is numerically evaluated with Gaussian quadrature using $\xi^{CC}$ and
$\xi^{G}$ both defined on the Gaussian grid.

\begin{figure}[htbp]
\centering
 \includegraphics[height=\linewidth,keepaspectratio,angle=90,bb=80 22 612 770]{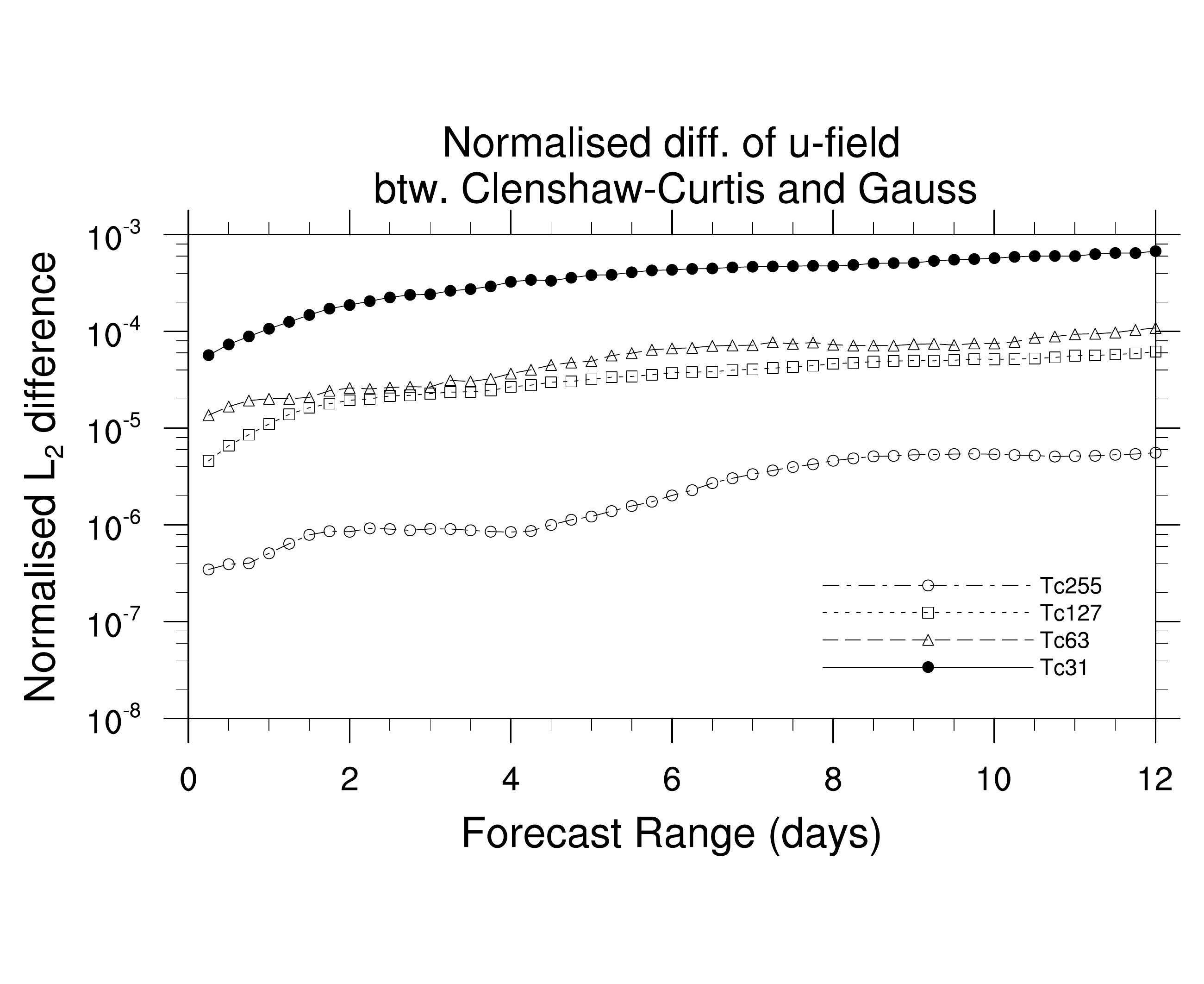}
 \caption{Normalised $L_2$-difference between Clenshaw-Curtis and
 Gaussian integration results for the zonal wind field plotted as a
 function of integration length. Note the log scale in the $y$-axis.}
 \label{fig:l2norm-u-testcase5}
\end{figure}

 The normalised $L_2$-differences of the zonal wind fields (Figure
\ref{fig:l2norm-u-testcase5}) are generally very small. They grow as the
integration length gets longer but seem to saturate by $\sim$ 10-day
integration. Higher resolutions result in smaller differences,
presumably due to reduced interpolation errors in semi-Lagrangian
advection. At saturation, the normalised differences are $\sim 10^{-3}$
with Tc31 resolution and $\sim 10^{-5}$ with Tc255 resolution, which are
in practice negligibly small. Similar results were also observed for $v$
and $h$ fields (not shown).

\subsection{Results: Barotropic jet instability}
\label{sec:results-galewsky}

\begin{figure}[htbp]
\centering 
 \includegraphics[height=\linewidth,keepaspectratio,bb=13 13 598 588,angle=90]{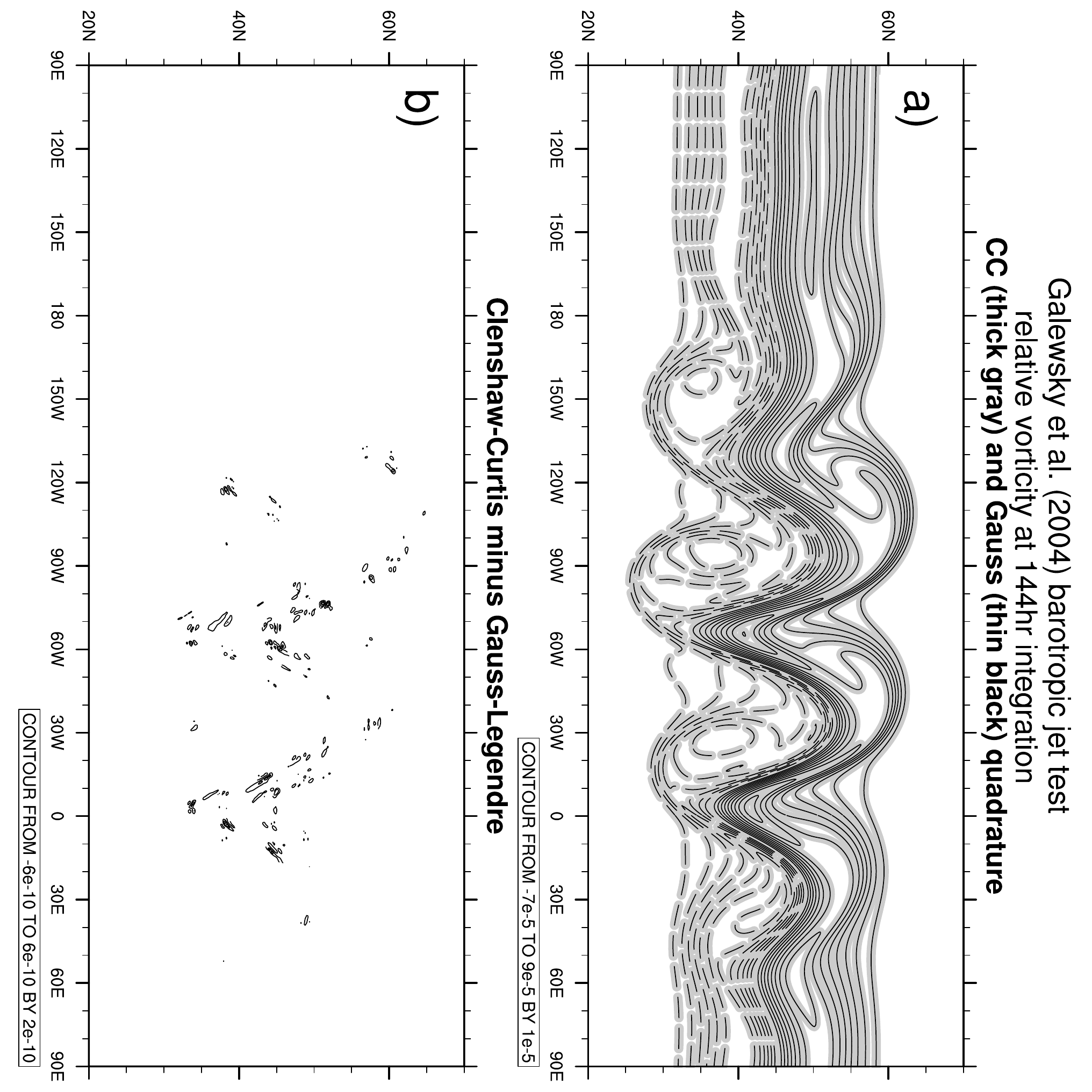}
 \caption{(a) Snapshot of relative vorticity field at day 6 from the barotropic jet
 instability test case produced with Clenshaw-Curtis (thick gray lines)
 and Gaussian (thin black lines) versions of the model at Tc479 ($\Delta x \sim 20$ km)
 resolution, and (b) the difference between the two. Contour intervals
 are $10^{-5}$ s in (a) and $2\times 10^{-10}$ s in (b).  Positive and
 negative values are drawn, respectively, with solid and dashed
 contours, and the zero contours are omitted. Note that in (a) the two sets
 of contour lines are completely superimposed.}
 \label{fig:galewsky}
\end{figure}

With sharp frontal structures present in the solution, in the barotropic
jet instability test, one may expect to see a larger discrepancy between
the two versions of the model with different quadrature rules and the
associated grids. We observed, however, that the two versions
yielded results that are visually indistinguishable (Figure
\ref{fig:galewsky}a), even at integration length as long as 144 hours (6
days) by which time multiple rolled-up vortical structures develop. The
difference between the two can be found mainly in the frontal regions
with strong vorticity gradients (Figure \ref{fig:galewsky}b), but the
differences are smaller than the raw values by more than five orders of
magnitude (note the different contour intervals in panels (a) and (b) of
Figure \ref{fig:galewsky}). The normalised $L_2$-difference for any of
$u, v$ and $h$ field (not shown) were below $10^{-7}$, again indicating
that the discrepancies arising from use of different quadrature rules
and the associated grids are negligible in practice.

\section{Implementation to a hydrostatic primitive equations model}

\subsection{The model}
Clenshaw-Curtis quadrature is implemented to the dry dynamical core of
JMA-GSM. It is an HPE model with Semi-Implicit Semi-Lagrangian (SISL)
time discretisation. The horizontal discretisation is spectral with
spherical harmonics as the basis functions, and the vertical
discretisation employs finite differencing on $\sigma-p$ hybrid
coordinate \citep{sb81} with 100 layers extending from the surface up to
0.01 hPa. Detailed description of the model's dynamical core can be
found in Section 3.1 of \cite{yukimoto11} and Section 3.2.2 of
\cite{jma-outline13}.

As with most other SISL spectral models, by
default JMA-GSM uses linear truncation. Since Clenshaw-Curtis quadrature
requires cubic (or higher) grid truncation, in this study
JMA-GSM is modified to allow for cubic truncation. All the results
shown in this section are produced with Tc159 horizontal resolution ($\Delta x\sim
60$ km at the Equator) and a time step $\Delta t= 1200$ s.

\subsection{Test case specification}

To assess any discrepancies in the forecasts that result from the
 use of Clenshaw-Curtis quadrature instead of the standard Gaussian
 quadrature, we perform the standardised test case proposed by
 \cite{jw06} and adopted by numerous studies that implement new schemes
 to global atmospheric dynamical cores. The Jablonowski-Williamson test
 case comprises two parts. The first part, the steady-state test,
 initialises the model with an analytic, baroclinically-unstable
 steady-state solution to the HPE, and tests the model's ability to
 maintain this steady-state initial condition. The second part, the
 baroclinic wave instability test, initialises the model with the same
 steady-sate solution but with a localised small-amplitude perturbation
 in the zonal wind field superimposed on the westerly jet axis located
 in the Northern Hemisphere mid-latitude. The small perturbation
 eventually develops into a train of unstable baroclinic waves that
 experiences nonlinear break-down by $\sim$ day 9. This second test
 allows us to examine the model's behaviour in a situation akin to a
 typical mid-latitude synoptic condition with sharp fronts and
 fine-scale eddies.

\subsection{Results}

\subsubsection{Steady-state test}
We first examine the impact of quadrature choice on the model's ability
to maintain the steady-state solution by comparing, between the
Clenshaw-Curtis and Gaussian versions of the model, the two measures of
steady-state maintenance proposed by \cite{jw06}. The first measure,
denoted as $l_2\left(u(t)-\bar{u}(t)\right)$ and defined as the square
root of the three-dimensional integral of zonally asymmetric part of the
zonal wind $u$ field (see Eq. (14) of \cite{jw06}), quantifies to what
extent the model is able to keep the zonally symmetric structure of the
steady initial state. The second measure, denoted as
$l_2\left(\bar{u}(t)-\bar{u}(t=0)\right)$ and defined as the square root
of the three-dimensional integral of the departure of the zonal
component $\bar{u}(t)$ of the zonal wind at time $t$ from that at the
initial time (see Eq. (15) of \cite{jw06}), complements the first
measure and quantifies how much the zonal mean of $u$ field deviates
from the initial zonally symmetric steady state over the course of
integration.

\begin{figure}[htbp]
\centering
 \includegraphics[width=\linewidth,bb=0 0 360 302]{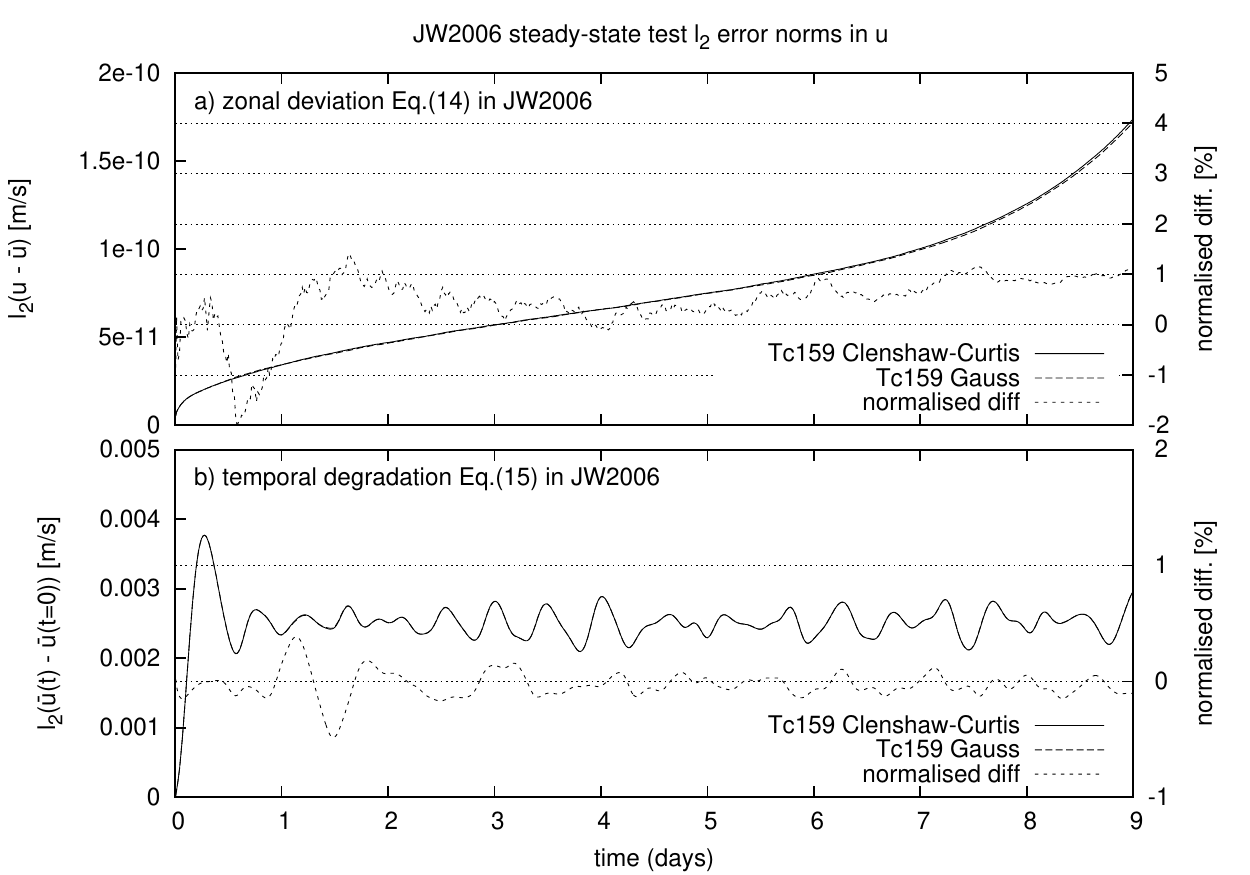}
 \caption{$l_2$ measures of steady-state maintenance in the steady-state
 test of \cite{jw06} computed for (solid) Clenshaw-Curtis and (dashed)
 Gaussian versions of the dynamical core of JMA-GSM. (a) the zonal
 symmetry measure $l_2\left(u(t)-\bar{u}(t)\right)$. (b) the temporal
 degradation of the zonal mean zonal wind
 $l_2\left(\bar{u}(t)-\bar{u}(t=0)\right)$. Note that the solid and
 dashed lines are almost superposed both in (a) and (b). The dotted
 lines (to be read on the right axis) represent the relative
 difference between the Clenshaw-Curtis and Gaussian results normalised
 by the values of Gaussian results.}
 \label{fig:l2norm-jw}
\end{figure}

As shown in Figure \ref{fig:l2norm-jw}, the Clenshaw-Curtis and Gaussian
versions of the model result in virtually identical $l_2$ error
measures. The relative difference between the two versions of the model
(defined as the difference normalised by the error of the Gaussian
version; drawn with dotted lines and to be read on the right axis) is
at most 2\% for the first measure (Figure \ref{fig:l2norm-jw}a) and
is well below $\sim $1 \% for the second measure (Figure
\ref{fig:l2norm-jw}b), indicating that the choice of the quadrature rule and
the associated grid does not affect the model's ability to maintain the
steady-state initial condition.

\subsubsection{Baroclinic instability wave test}

\begin{figure}[htbp]
\centering 
 \includegraphics[height=\linewidth,keepaspectratio,bb=13 13 598 588,angle=90]{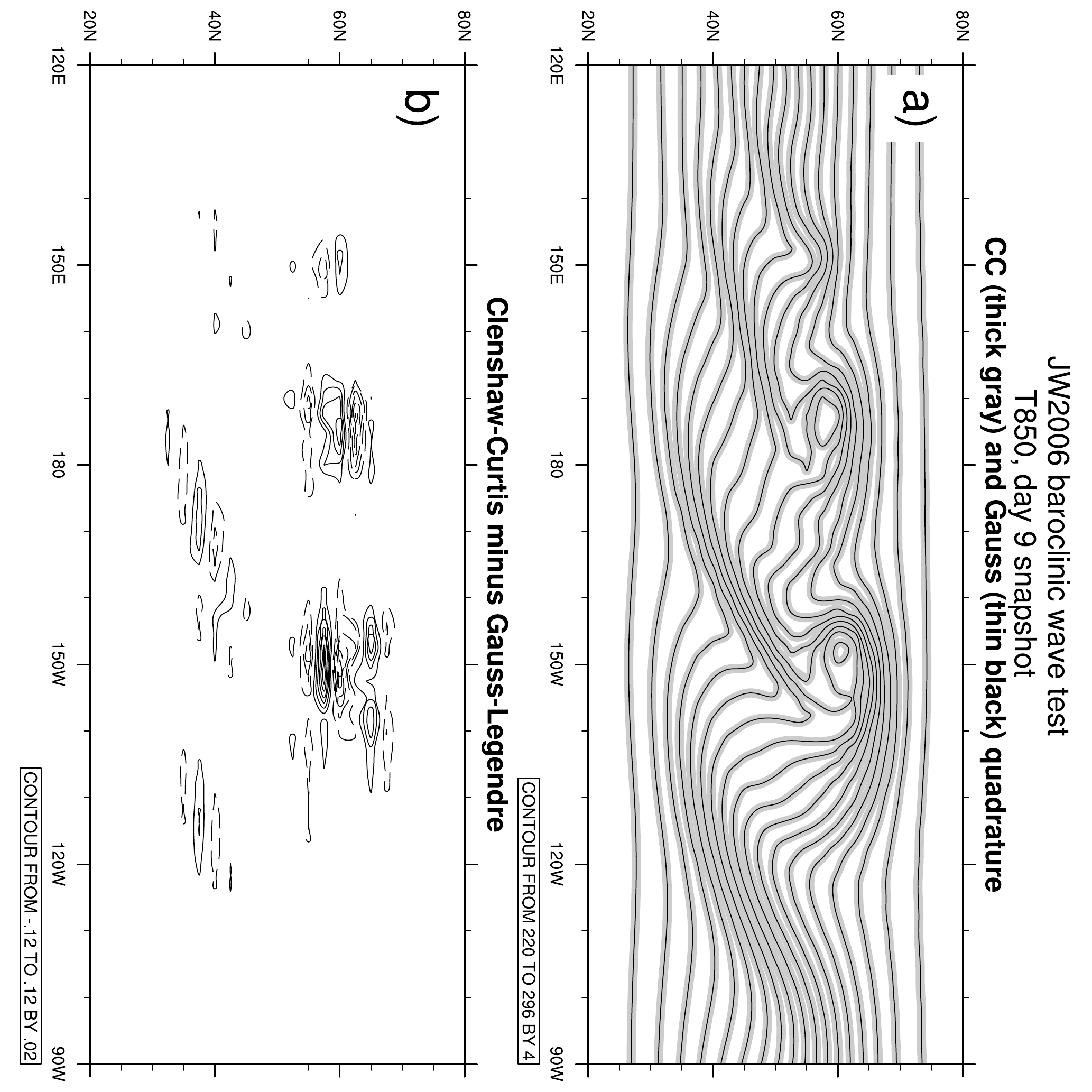}
 \caption{(a) Snapshot of 850-hPa temperature at day 9 from the
 baroclinic instability wave test case produced with Clenshaw-Curtis
 (thick gray lines) and Gaussian (thin black lines) versions of JMA-GSM
 dynamical core run at Tc159 resolution, and (b) the difference between
 the two. Contour intervals are 4 K in (a) and 0.02 K in (b). In (b),
 positive and negative values are drawn, respectively, with solid and
 dashed contours, and the zero contours are omitted. Note that in (a) the
 two sets of contour lines are completely superimposed.}
 \label{fig:jw-instability}
\end{figure}

As in the barotropic jet instability test for SWE model (section
\ref{sec:results-galewsky}, Figure \ref{fig:galewsky}), with sharp
fronts present in the solution, we can expect the baroclinic instability
wave test to better reveal discrepancies (if any) in the numerical results
arising from the use of different quadrature rules and the associated
grids. Just like in section \ref{sec:results-galewsky}, however,
again we confirmed a high degree of agreement between the solutions from
Clenshaw-Curtis and Gaussian versions of the model. As an example, we
show the snapshots of 850 hPa temperature field at day-9 integration in
Figure \ref{fig:jw-instability}a. The results from Clenshaw-Curtis
version (thick gray contours) and Gaussian version (thin black contours)
are indistinguishable on this plot, and the difference between the two
(Figure \ref{fig:jw-instability}a) exhibit noisy patterns localised
along the fronts and with very small amplitude (at most 0.13 K),
indicating that no systematic difference results from the use of
different quadrature rules.

\section{Conclusion and future directions}

Traditional global spectral atmospheric models adopt Gaussian
quadrature, and the non-nested property of its associated irregular
latitudinal quadrature nodes (the Gaussian grid) makes it difficult if
not impracticable to apply multigrid approaches to global spectral
models.  To facilitate straightforward use of a multigrid method in a
global spectral model in the future, in this study we proposed to adopt
Clenshaw-Curtis quadrature in evaluating direct (grid-to-wave) Legendre
transform. With Clenshaw-Curtis quadrature, the latitudinal grids (or
quadrature nodes) are equispaced, so that the model grid for an
arbitrary horizontal resolution includes the model grid for half its
horizontal resolution as its complete subset, allowing for
implementation of a multigrid method without a need for any off-grid
interpolation.

Theoretical consideration shows, and numerical computation demonstrated,
that Clenshaw-Curtis quadrature guarantees exact (up to machine
precision) spherical harmonics transforms if it is used with cubic (or
higher-order) grid truncation. One may argue that having to move from
linear or quadratic truncation, which are the norm for the current
semi-Lagrangian or Eulerian spectral models, respectively, is a serious
limitation, but that is not the case for high-resolution modelling
since, as shown by recent studies at ECMWF \citep{wedi14}, higher-order
truncation is anyhow necessary to suppress aliasing errors and the
resultant spectral blocking at a resolution as high as $\Delta x <$ 10
km.

Comparison of the Clenshaw-Curtis and Gaussian versions of the
shallow-water model and dry hydrostatic primitive equations model, both
with cubic grid and with semi-Lagrangian advection scheme, performed
under the framework of idealised standard test cases, demonstrated that
Clenshaw-Curtis quadrature gives model predictions that are almost
identical to those obtained with the standard Gaussian quadrature. 

\begin{figure}[htbp]
\centering 
 \includegraphics[width=\linewidth]{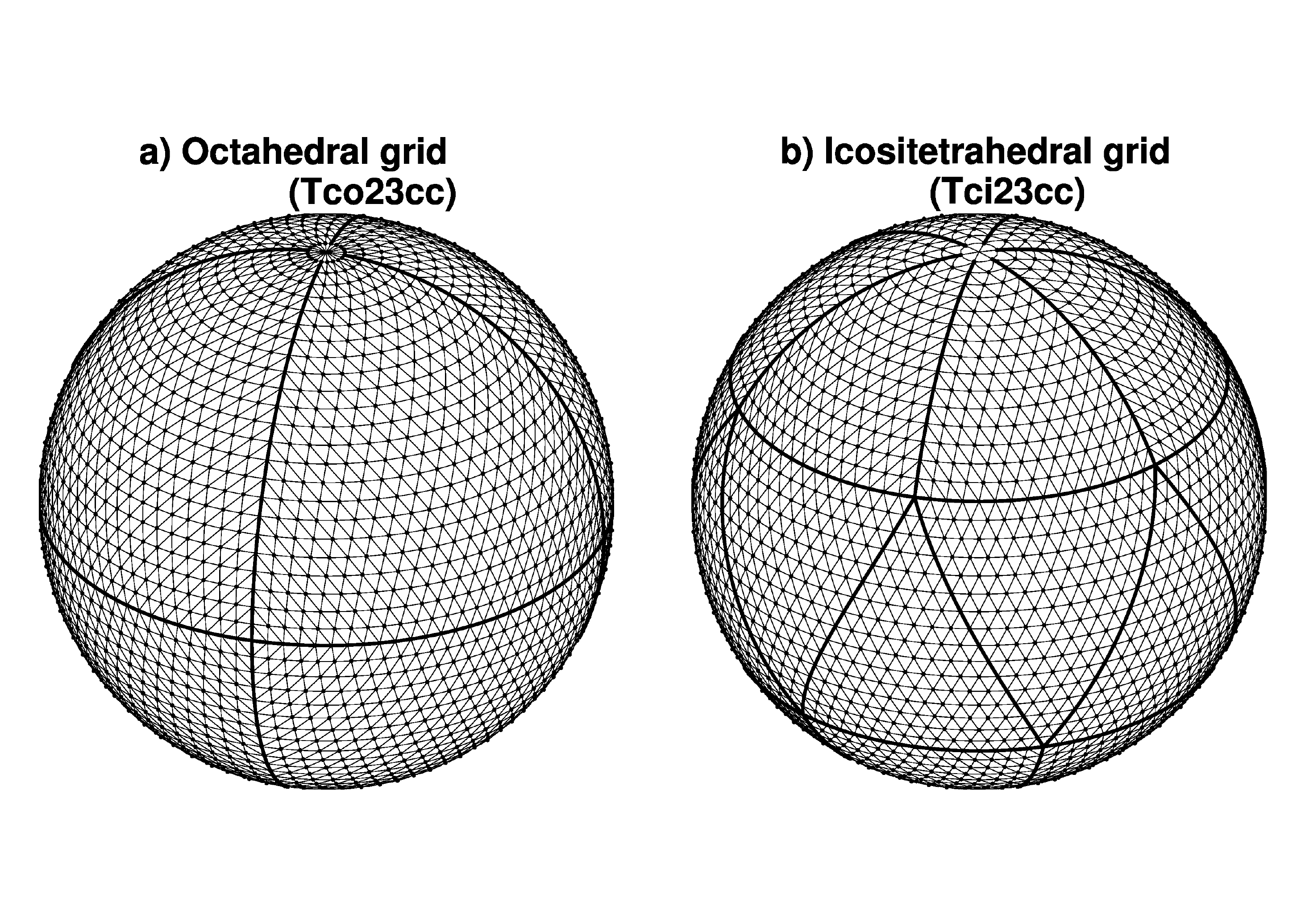}
 \caption{ Cubic Clenshaw-Curtis (a) octahedral and (b) icositetrahedral
 grid for Tc23 resolution (which corresponds to 3.75\textdegree grid spacing in
 the meridional direction).}
 \label{fig:cc-grids}
\end{figure}

Our ultimate goal is to ensure computational efficiency of a global
spectral non-hydrostatic atmospheric model by enabling a multigrid
approach, as outlined in Appendix C, and this study serves as a first
step in this direction. The success (or not) of such an approach is not
clear, particularly because the future evolution of the high-performance
computing (HPC) architecture is still in a state of flux. It has long
been predicted that spectral models will face a serious challenge on a
massively parallel architecture because their current algorithms require
global inter-node communications on every time step. To maintain
admissible scalability on future architectures, therefore, global models
would have to abandon, or at least reduce reliance on, spectral
transforms. In this line, ECMWF has recently introduced a structured
grid system, called `cubic octahedral reduced Gaussian grid' that
assures exact spectral transforms via Gaussian quadrature
\citep{malardel16} while at the same time allowing for straightforward
grid-based discretisation of horizontal derivatives that only requires
neighbouring-node communications \citep{smolarkiewicz16}, whereby paving
a path for a smooth transition from spectral to grid-based modelling (or
hybridisation of the two methods). Their approach of allowing both
spectral and grid-based discretisations on a single grid system is
ingenious and appealing, and we assert that Clenshaw-Curtis quadrature,
with its equispaced latitude grids, can be combined to yield cubic
octahedral Clenshaw-Curtis grid (Figure \ref{fig:cc-grids}a).

Compared to the Gaussian octahedral grid, the Clenshaw-Curtis octahedral
grid has the advantage of allowing for a straightforward use of
multigrid in gridpoint-space solution of elliptic equations that result from
a (semi-)implicit time stepping. Moreover, it covers the globe entirely
with triangular meshes (Figure \ref{fig:cc-grids}a), unlike the Gaussian
counterpart which has rectangular cells along the Equator. Its
latitudinally-uniform grid spacing may also help to make the derivative
stencils somewhat simpler. 

\begin{figure}[htbp]
\centering 
 \includegraphics[width=\linewidth,bb=0 0 292 169]{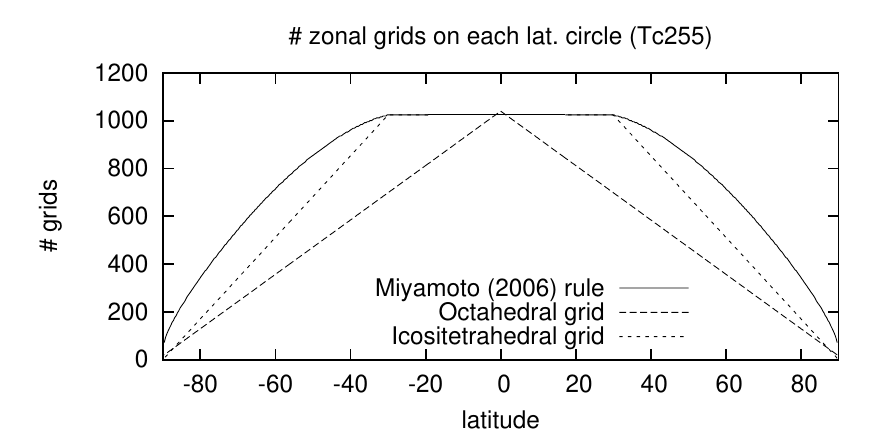}
 \caption{The number of gridpoints on each latitude circle for Tc255
 resolution decided by (solid line) the grid reduction rule proposed by
 \cite{miyamoto06} (see the text for detail), (dashed line) octahedral
 grid alignment, and (dotted line) icositetrahedral grid alignment.}
 \label{fig:numgrid-vs-lat}
\end{figure}

Alternative grid alignments based on reduced
Clenshaw-Curtis grid are also possible that cover the globe with
triangular meshes. Figure \ref{fig:cc-grids}b shows one such
possibility, which may be called cubic icositetrahedral (24-face
polyhedral) Clenshaw-Curtis grid, where the globe is divided into 24
triangular panels of equal area, 6 covering the Northern Hemisphere
extratropics (from 90\textdegree N down to 30\textdegree N), 12 covering
the tropical 30\textdegree N--30\textdegree S belt, and the remaining 6
covering the rest (from 90\textdegree S up to 30\textdegree S). Compared
to the octahedral grid, having to shift the phase in zonal Fourier
transforms in the tropical belt introduces additional complexity in the
code, but the icositetrahedral grid gives more uniform grid spacing
across latitudes and is more compatible (Figure
\ref{fig:numgrid-vs-lat}) with the grid reduction rule of
\cite{miyamoto06} which guarantees numerically exact spherical harmonics
transform by requiring, for each latitude $\phi_j$, that there be at
least $4M_j+1$ longitudinal points (with cubic truncation) where 
$M_j$ is defined as the smallest zonal wavenumber for which
$\left|\tilde{P}_n^m\left(\sin\phi_j\right)\right|< 10^{-16} $ ($\sim$
double precision machine epsilon) holds for any $0\leq n\leq N$ and
$|m|\leq M_j$. The effectiveness in terms of accuracy, computational
efficiency on a massively parallel machine architecture, and ease of
transition from the current spectral model, of using octahedral and/or
icositetrahedral Clenshaw-Curtis grids, along with gridpoint-space-based
multigrid methods, will be explored in our future project.

\section*{Acknowledgements}
 The authors thank Prof. Keiichi Ishioka (Kyoto University) and
Hiromasa Yoshimura (MRI-JMA) for their useful comments on the early
version of the manuscript and two anonymous reviewers for their numerous
constructive suggestions that have led to substantial improvement of the
manuscript.

\section*{Appendix A: Derivation of the Clenshaw-Curtis quadrature weights $w^{CC}_j$} 
\setcounter{equation}{0}
\renewcommand{\theequation}{A\arabic{equation}}

Let $g(\cos\theta)$ be the integrand as defined in (\ref{eq:defg}) and
assume that $g(\cos\theta)$ can be expanded by the Chebyshev polynomials
of the second kind of degrees up to $J-1$, giving
\begin{eqnarray}
 g(\cos\theta) \sin\theta= \sum_{l=0}^{J-1}a_l U_l\left(\cos\theta\right)\sin\theta = \sum_{p=1}^{J}a_{p-1}\sin{p\theta}
\end{eqnarray}
The FFT-based derivation of the quadrature obtains the expansion
coefficients $a_l$ by performing DST on $g(\cos{\theta})\sin\theta$ and
then analytically integrates each summand in the rightmost-hand side of
(A1). The alternative derivation by \cite{boyd87} expands
$g(\cos\theta)\sin\theta$ into a linear combination of ``trigonometric
cardinal functions'' $\left\{C_j(\theta)\right\}_{j=1}^J$ rather than
the sine series:
\begin{eqnarray}
 g(\cos\theta) \sin\theta = \sum_{j=1}^J\left\{ g(\cos\theta_j) \sin\theta_j \right\} C_j(\theta)
\end{eqnarray}
where $\left\{C_j(\theta)\right\}_{j=1}^J$ are the transformed basis of
the space spanned by the sine series
$\left\{\sin{p\theta}\right\}_{p=1}^{J}$ that satisfies
\begin{eqnarray}
 C_j(\theta_i) = \delta_{ij} 
\end{eqnarray}
for any $i,j\in\left\{1,2,\cdots,J\right\}$ with the nodes $\theta_j$
chosen as in (\ref{eq:cc-points}). As shown in the Appendix of
\cite{boyd87}, by exploiting the discrete orthonormality of
$\left\{\sin{p\theta}\right\}_{p=1}^{J}$ on the points in
(\ref{eq:cc-points}), the cardinal functions $C_j(\theta)$ can be
expressed as
\begin{eqnarray}
 C_j(\theta) &=& \frac{2}{J+1}\sum_{p=1}^J\sin{p\theta_j}\sin{p\theta}.
\end{eqnarray}
The explicit expression of the quadrature weights $w^{CC}_j$ as shown
in (\ref{eq:cc-wgt}) follows from plugging (A4) into (A2) and then
carrying out the integration from $\theta=0$ to $\theta=\pi$.  \hfill
$\blacksquare$

\section*{Appendix B: A comment on parity exploitation in spherical harmonics transforms}
\setcounter{equation}{0}
\renewcommand{\theequation}{B\arabic{equation}}

The associated Legendre functions $\tilde{P}_n^m(\sin\phi)$ are all
either symmetric (if $n\equiv m \mod 2$) or antisymmetric (otherwise)
across the Equator. This parity property, along with the hemispheric
symmetry of the Gaussian latitudes and the Gaussian weights $w^G_j$, can
be exploited to halve the computational cost of preparing
$\tilde{P}_n^m(\sin\phi_j)$ and $w^G_j$ (since they are only necessary
for $j\leq J/2$) and of performing direct or inverse associated Legendre
transforms, while at the same time preserving the hemispheric symmetry
of the solutions. For example, the direct transform
(\ref{eq:gauss-quad}) can be rewritten as
\begin{eqnarray}
 X_n^m = \sum_{j=1}^{J/2} X^m_S(\phi_j)\tilde{P}_n^m(\sin\phi_j)w^G_j\quad n\equiv m \mod 2\\
 X_n^m = \sum_{j=1}^{J/2} X^m_A(\phi_j)\tilde{P}_n^m(\sin\phi_j)w^G_j\quad n\not\equiv m \mod 2
\end{eqnarray}
where the decomposition of $X^m(\phi_j), j=1,\cdots,J$ into its symmetric and
antisymmetric components $X^m_S(\phi_j):=X^m(\phi_j)+X^m(\phi_{J+1-j})$
and $X^m_A(\phi_j):=X^m(\phi_j)-X^m(\phi_{J+1-j}), j=1,\cdots,J/2$,
which takes only $O(J)$ operations, is assumed to be performed prior to
the transform. The inverse transform (\ref{eq:iLT}) can be similarly
economised. 

The same technique can be readily applied to the Clenshaw-Curtis version
of the associated Legendre transforms because, as in the Gaussian case,
Clenshaw-Curtis latitudes and weights are symmetric across the
Equator. The only caveat to note here is that, unlike the Gaussian,
Clenshaw-Curtis latitudes include the Equator, resulting in odd numbers
of latitudinal gridpoints in total. Hence, the parity-exploited version of
the direct transform becomes:
\begin{eqnarray}
 X_n^m = \sum_{j=1}^{(J+1)/2} X^m_S(\phi_j)\tilde{P}_n^m(\sin\phi_j)w^{CC}_j\quad n\equiv m \mod 2\\
 X_n^m = \sum_{j=1}^{(J+1)/2} X^m_A(\phi_j)\tilde{P}_n^m(\sin\phi_j)w^{CC}_j\quad n\not\equiv m \mod 2
\end{eqnarray}
where $X^m_S(\phi_j)$ and $X^m_A(\phi_j)$ are defined as in (B1,B2)
except for $X^m_S(\phi_{(J+1)/2}):=X^m(\phi_{(J+1)/2})$ 
and $X^m_A(\phi_{(J+1)/2}):=0$, and likewise for the inverse transform.

As stated above, parity exploitation not only saves computational costs
but also helps to preserve the symmetry of solutions over the two
hemispheres. This can be confirmed, for instance, by meridionally
flipping the initial condition, running the model, and then meridionally
flipping the solution again and checking if the solutions with and without
flipping operations agree with each other. We have confirmed that this
holds for our implementation using the Galewsky test described in
section \ref{sec:results-galewsky} (not shown).

 \section*{Appendix C: Sketch on how Clenshaw-Curtis
 quadrature and grid will simplify multigrid implementation} 
\setcounter{equation}{0}
\renewcommand{\theequation}{C\arabic{equation}}

To clarify the motivation behind introducing the nestable
Clenshaw-Curtis grid in a global spectral atmospheric model, in this
appendix we briefly describe how the proposed grid and quadrature will
help to implement multigrid methods in a global spectral atmospheric
model. Readers unfamiliar with multigrid nomenclature are invited to
refer to, for example, the review by \cite{fulton86}. Actually
implementing multigrid methods to a global spectral model is beyond the
scope of this paper and is planned to be addressed in our future work.

Let us denote the governing equations of the non-hydrostatic atmospheric dynamics
symbolically as
\newcommand{\calM}{\mathcal{M}}
\newcommand{\calL}{\mathcal{L^*}}
\newcommand{\calN}{\mathcal{N}}
\newcommand{\calI}{\mathcal{I}}
\begin{eqnarray}
 \frac{DX}{Dt}=\calM(X) \label{eq:gov-eq}
\end{eqnarray}
where $X$ is the state vector and $\calM$ is the nonlinear tendency
operator. As discussed in \cite{benard04}, there are two semi-implicit approaches,
both iterative, to stably integrate \ref{eq:gov-eq}) with a long time step.

In the first, non-constant-coefficient SISL approach, we divide $\calM(X)$ into slow and fast
components
\begin{eqnarray}
 \calM(X)=\calM^\mathrm{slow}(X)+\calM^\mathrm{fast}(X) \label{eq:fast-slow}
\end{eqnarray}
and the fast part $\calM^\mathrm{fast}(X)$ is further partitioned into
its tangent-linearisation $\calL(X-X^*)$ around some reference state $X^*$ and the
residual $\calM^\mathrm{fast}(X)-\calL(X-X^*)$, yielding:
\begin{eqnarray}
 \frac{DX}{Dt}=\calN(X)+\calL{X}, \label{eq:nonlin-lin}
\end{eqnarray}
where the operator $\calN$ is defined so that (\ref{eq:gov-eq}) combined with (\ref{eq:fast-slow})
matches (\ref{eq:nonlin-lin}), which is then descretised in time to give
\begin{eqnarray}
 \frac{X_A^+ - X_D^0}{\Delta t} =
  \overline{\calN(X)}^{(+1/2)}+\frac{1}{2}\left(\calL{X_A^+}+\calL{X_D^0}\right) \label{eq:2tl-sisl}
\end{eqnarray}
where subscripts $A$ and $D$ denote values at the arrival and departure
points, respectively, superscripts $+$ and $0$ denote values at the
future and current time steps, respectively, and the overline
$\overline{(\cdot)}$with superscript $(+1/2)$ denotes mean over the
trajectory from current time at departure point to future time at
arrival point, approximated by some extrapolation scheme such as SETTLS
\citep{hortal02} using the past and the current time step. Arranging all
the unknown future state on the left hand side and all the known
quantities on the right, (\ref{eq:2tl-sisl}) results in the following
linear systems of equations that can be arranged into a Helmholtz-type
elliptic equation
\begin{eqnarray}
\left(\calI - \Delta t/2\calL\right)X_A^+ = F \label{eq:elliptic}
\end{eqnarray}
where $\calI$ is the identity operator.  In contrast to a hydrostatic
system where choosing the reference state $X^*$ to be horizontally
homogeneous, which renders (\ref{eq:elliptic}) a constant-coefficient
elliptic equation that can be efficiently solved in spectral space,
allows sufficient stability, non-hydrostatic dynamics requires the
reference state $X^*$ to be inhomogeneous to bring $\calL$ closer to
$\calM^\mathrm{fast}$ to achieve reasonable stability, rendering
(\ref{eq:elliptic}) a non-constant-coefficient elliptic equation that
needs to be solved iteratively. 

An alternative, `more implicit' approach stabilises the integration by
approximately solving the fully implicit time discretisation of (\ref{eq:gov-eq}):
\begin{eqnarray}
 \frac{X_A^+ - X_D^0}{\Delta t} = \frac{1}{2}\left(\calM(X_A^+)+\calM(M_D^0)\right) \label{eq:full-imp}
\end{eqnarray}
with some generic nonlinear elliptic solver or with an iterative
centred-implicit (ICI) scheme \citep{benard04} that allows to exploit
the spectral constant-coefficient semi-implicit solver.

Either approach results in iterative solution of a linear or nonlinear
elliptic equation, for which multigrid is an effective approach. The
nested property of Clenshaw-Curtis grid (section \ref{sec:lowres-sht}),
while not necessarily reduces computational cost in terms of operation
count, will allow an easier implementation (with less coding effort and
possibly less inter-node communications) of multigrid operations than the
non-nested Gaussian grid would require. Restriction (grid transfer from
higher- to lower-resolution) and prolongation (from lower- to
higher-resolution) in spectral space are trivial since the
lower-resolution spectral space is a proper subset of the
higher-resolution spectral space.  Restriction in grid space can be
achieved by simple injection (or thinning; skipping every other grid)
thanks to the nested property; note that non-nested grid like the
Gaussian would necessitate off-grid interpolation, requiring some extra
coding effort and possibly additional inter-node communication in a
parallelised environment. Restriction from the higher-resolution grid
space to the lower-resolution spectral space can be economically
performed by the mechanism discussed in section \ref{sec:lowres-sht},
unlike in the Gaussian case where a costly high-resolution forward
transform would be necessary. Prolongation in grid space can be achieved
cheaply and locally ({\it i.e.,} without a need for inter-node
communication) by simple linear or spline interpolation in situations
where spectral accuracy is not desired.

Using the economical grid transfer operations discussed above as
building blocks, any multigrid solvers should be able to be constructed
straightforwardly. As an illustrative example, consider the simplest-possible
multigrid scheme, a V-cycle with only two grids (with Tc479 finer grid
and Tc239 coaser grid, for example) applied to solving the non-constant
coefficient elliptic problem (\ref{eq:elliptic}). The algorithm proceeds as follows:
\begin{enumerate}
 \item At Tc479 resolution, approximately solve (\ref{eq:elliptic}) by
       some elliptic solver, e.g., iterative relaxation or Krylov
       subspace scheme in grid space starting from, e.g., the spectrally
       computed solution of the constant-coefficient approximation of
       (\ref{eq:elliptic}), with the horizontal derivatives evaluated
       pseudo-spectrally. At this point iteration does not have to be
       repeated until convergence, but the high wavenumber components
       should reasonably converge. Denote the current approximation by
       $X_{1}^\mathrm{Tc479}$ and compute the residual
       $R^\mathrm{Tc479}:=(\calI-\Delta
       t/2\calL)X_{1}^\mathrm{Tc479}-F$.
 \item Restrict $R^\mathrm{Tc479}$ into Tc239 space to obtain
       $R^\mathrm{Tc239}$ and solve the residual equation
       $(\calI-\Delta t/2\calL)Y= -R^\mathrm{Tc239}$ for $Y$. This can
       be done by the same method as in step 1. but at the lower
       resolution. The resultant $Y$ is a lower-resolution approximation
       to the error $X_A^+ - X_{1}^\mathrm{Tc479}$.
 \item Prolong $Y$ to Tc479 resolution, then add  to
       $X_{1}^\mathrm{Tc479}$ to yield the improved approximate solution
       $X_{2}^\mathrm{Tc479}$, and
       repeat the relaxation as in step 1., starting from
       $X_{2}^\mathrm{Tc479}$ as the first guess, with a tighter
       convergence criterion to ensure further improved approximation.
\end{enumerate}
The `more implicit' solver that approximately solves (\ref{eq:full-imp})
should be similarly accelerated by a multigrid approach by using, e.g.,
the nonlinear multigrid full approximation scheme (FAS).

Finally, we postulate that employing the pseudo-spectral multigrid approach as
outlined above will foster smooth and gradual transition from spectral
modelling to grid-based (or grid/spector hybrid) modelling since, in the
above-outlined framework, the grid-space representation of the
horizontal derivatives evaluated by the pseudo-spectral method can be
readily replaced by local horizontal derivatives evaluated by some
grid-based scheme such as finite difference, finite volume, or
finite/spectral element. Given that grid-based elliptic solvers tend to
be less efficient at larger scale, a grid/spector hybrid approach, where
a grid-based multigrid method with shallow layers is combined with a spectral
elliptic solver used only at the coarsest grid with moderate resolution,
seems a reasonable strategy that compromises the need to avoid global
inter-node communications and to maintain acceptable accuracy and
convergence rate.


\newcommand{\QJRMS}{{\it Q.  J.  R. Meteorol. Soc.}\ }
\newcommand{\MWR}{{\it Mon. Wea. Rev.}\ } \newcommand{\BAMS}{{\it
Bull. Am. Meteor. Soc.}\ } \newcommand{\JAS}{{\it J. Atmos. Sci.}\ }
\newcommand{\JGRA}{{\it J. Geophys. Res.-Atmos.}\ }
\newcommand{\JGRO}{{\it J. Geophys. Res.-Oceans.}\ }
\newcommand{\JCP}{{\it J. Comp. Phys.}\ } \newcommand{\bluebook}{{\it
CAS/JSC WGNE Res. Activ. Atmos. Oceanic Modell.}\ }

\end{document}